\documentclass{aa}  

\usepackage{graphicx}
\usepackage{xcolor}
\usepackage{txfonts}
\usepackage{lipsum}
\usepackage{subcaption}         
\usepackage{lscape}             
\usepackage{placeins}           
                                
\newcommand{\vel}{\,km s$^{-1}$}

\newcommand{\msun}{\,h^{-1}{\rm M_\odot}}
\newcommand{\OG}{\textsc{OpenGadget3}}
\newcommand{\SF}{\textsc{SubFind}}

\newcommand{\mincir}{\raise
  -2.truept\hbox{\rlap{\hbox{$\sim$}}\raise5.truept \hbox{$<$}\ }}
\newcommand{\magcir}{\raise
  -2.truept\hbox{\rlap{\hbox{$\sim$}}\raise5.truept \hbox{$>$}\ }}

\begin{document}

   \title{{\texttt{Dianoga}} simulations of galaxy clusters and groups}

   \subtitle{Properties of the baryonic components}

%
%
%

\author
{Stefano Borgani\inst{1,2,3,4,5}\corrauth{sborgani@units.it}
\and Alice Damiano\inst{1,2,4,5}
\and Elena Rasia\inst{2,3,6}
\and Tiago Castro\inst{2,8}
\and Michela Esposito\inst{2}
\and Ilaria Marini\inst{8,9,10}
\and Giuseppe Murante\inst{2,3,5}
\and Milena Valentini\inst{1,2,3,4,5}
\and Veronica Biffi\inst{2,3}
\and Klaus Dolag\inst{10,11,9}
\and Gian Luigi Granato\inst{2,12}
\and Antonio Ragagnin\inst{2}
\and Cinthia Ragone-Figueroa\inst{12,2}
\and Alex Saro\inst{1,2,3,4,5}
\and Luca Tornatore\inst{2,5}
}

\institute{Dipartimento di Fisica dell'Universit\`a di Trieste, Sez. di Astronomia, via Tiepolo 11, I-34131 Trieste, Italy
\and INAF -- Osservatorio Astronomico di Trieste, via Tiepolo 11, I-34131, Trieste, Italy 
\and IFPU, Institute for Fundamental Physics of the Universe, Via Beirut 2, 34014 Trieste, Italy
\and INFN, Instituto Nazionale di Fisica Nucleare, Via Valerio 2, I-34127, Trieste, Italy
\and ICSC - Italian Research Center on High Performance Computing, Big Data and Quantum Computing
\and Department of Physics, University of Michigan, 450 Church Str., Ann Arbor, MI 48109, USA
\and Department of Mathematical Physics, Institute of Physics, University of São Paulo, R. do Matão 1371, 05508-090, São Paulo, SP, Brazil
\and European Southern Observatory, Karl Schwarzschildstrasse 2, 85748, Garching bei München, Germany
\and Excellence Cluster ORIGINS, Boltzmannstr. 2, D-85748 Garching bei M\"unchen, Germany
\and Universit\"ats-Sternwarte M\"unchen, Scheinerstr. 1, D-81679, M\"unchen, Germany
\and Max-Plank-Institut f\"ur Astrophysik, Karl-Schwarzschild Strasse 1, D-85740 Garching, Germany
\and Instituto de Astronom\'ia Te\'orica y Experimental (IATE), Consejo Nacional de Investigaciones Cient\'ificas y T\'ecnicas de la Rep\'ublica Argentina (CONICET), Universidad Nacional de C\'ordoba, Laprida 854, X5000BGR, C\'ordoba, Argentina
}



 
  \abstract
   {Reproducing the observed co-evolution between cluster galaxies, thermodynamical properties of the intra-cluster and intra-group media (ICM/IGM), and central supermassive black holes (SMBHs) remains a major challenge for cosmological hydrodynamical simulations.}
  {We introduce the {\tt Dianoga} set of cosmological simulations of galaxy clusters and groups, specifically aimed at studying the implication of this co-evolution on the implementation of AGN feedback and star formation.}
{Using the \OG\ code, we carry out simulations of 28 regions centred on massive galaxy clusters, and of a cosmological box. This generates a sample of 293 halos with $M_{200}\ge 1.5\times 10^{13}\mathrm{M_\odot}$. Parameters of AGN feedback in the reference implementation were minimally calibrated exclusively to match the local relation between SMBH masses and stellar masses of host galaxies. Simulations are compared to observed galaxy stellar mass function (GSMF), stellar mass fraction in clusters and groups, BCG masses, scaling relations between ICM/IGM properties and profiles of their thermodynamical properties. In the appendix, we show how results vary as we modify the reference feedback model in six alternative configurations.}
{Our reference model predicts a GSMF in general agreement with observations, albeit overestimated in the high end. BCG stellar masses and mass fractions are higher than observed in massive clusters, while being closer to observations for groups. Predicted properties of the ICM/IGM are in general agreement with observations, while the core regions simulated clusters have entropy and temperature profiles that are less "cool-cored" than observed. A comparison with other implementations of AGN feedback highlights that models including thermal evaporation of the sub-resolution interstellar medium succeed to bring BCG masses and stellar mass fractions closer to observation, and to increase cool-coreness of simulated clusters.}
{Our results demonstrate that the details of the interface between AGN energy injection and the sub-resolution interstellar medium model are at least as critical as the total feedback efficiency itself.}


   \maketitle
\nolinenumbers


\section{Introduction}
Since the very early days of computational cosmology, clusters of galaxies represented the target of choice to investigate the development of non-linear gravitational instability in the formation of cosmic structures \citep[e.g.][]{Peebles.1970,White.1976} and the hydrodynamical evolution of the cosmic baryons within the potential wells dominated by dark matter \citep[DM; e.g.][]{Evrard.1990,Navarro.etal.1995}. In fact, the widespread interest in clusters of galaxies has multiple reasons: they sit atop of the hierarchy of cosmic structures and, as such, they are privileged tracers of cosmic evolution; they are invaluable astrophysical laboratories, since they provide the most extreme ecosystems where galaxies, different phases of the intergalactic medium and super-massive black holes (SMBHs) co-evolve \citep[e.g.][]{gaspari.etal.2019}; they are unique places where to study the thermo- and chemo-dynamics of the hot intra-cluster and intra-group mediua (ICM/IGM; e.g. \citealt{voit2005}; \citealt{Biffi.etal.2018}); their number density and clustering as a function of mass and redshift are potentially highly sensitive cosmological probes \citep[e.g.][]{allen.etal.2011,kravtsov_borgani}. In this context, cosmological hydrodynamical simulations are instrumental to capture the complex interplay between cosmic evolution and galaxy formation processes, and in establishing the observational properties of galaxy clusters at different wavelengths \citep[e.g.][]{borgani_kravtsov2011}. 

While there is a long track record of successes of cosmological hydrodynamical simulations in reproducing the ICM thermodynamical properties outside the core cluster regions, where gravity is the driver of the evolution, the situation is more complicated in the core regions. In these regions, the possibility of carrying out spatially resolved spectroscopy with the Chandra and XMM-Newton satellites revealed since more than 20 years that the amount of gas cooling is much lower than predicted by the classical isobaric cooling model applied to the intense emissivity observed in relaxed clusters \citep[e.g.][]{Boehringer.etal.2002}. Consistently, the entropy level observed in such "cool core" clusters \citep[e.g.][]{Molendi2001,voit2005} highlights that gas with a formally short cooling time remains in a hot, X-ray emitting phase thanks to a gentle heating from a central AGN \citep[e.g.][]{McNamara2012}. At the same time, observational determinations of the star formation rate (SFR) measured at low redshift in Brightest Cluster Galaxies (BCGs; e.g. \citealt{McDonald.etal.2018}) is very low and consistent with the small amount of cooling gas, with only few remarkable exceptions of strongly star forming BCGs, such as the Phoenix cluster that reaches values of $\sim 10^3$~M$_\odot$/yr at redshift $z\simeq 0.6$ \citep[e.g.][]{Reefe.etal.2025}. The emerging observational picture thus demonstrates that the cooling-feedback loop at the centre of clusters should be determined by a self-regulated co-evolution of ICM, BCG and central SMBH \citep[e.g.][]{Gaspari2020,donahue_voit}. 

Not surprisingly, early studies of clusters through radiative simulations not including feedback from Active Galactic Nuclei (AGN) have failed in reproducing the observational properties of ICM in core regions \citep[e.g.][and references therein]{Borgani.etal.2004}. This "overcooling" problem, which resulted in an excessive production of stars at the centers of massive halos, highlighted the need for energetic AGN feedback. Over the last decade, several major simulation campaigns have sought to address the challenge of reproducing the cool core structure in simulations including AGN feedback, also with varying numerical techniques and sub-grid physics prescriptions.

In fact, with the inclusion of AGN feedback in  simulations \citep[e.g.][]{Springel.etal.2005,Sijacki.etal.2007}, the above tensions with observations have been partly alleviated, at least in the comparison with the observed scaling relations between X--ray luminosity and temperature of clusters and groups \citep[e.g.][]{Puchwein.etal.2008}. Over the last decade, different sets of simulations proved to be successful in reproducing the low-entropy level and stellar masses of the BCGs, using either zoom-in simulations, such as a lower-resolution version of the same {\tt Dianoga} set considered here \citep[e.g.][]{rasia.etal.2015,ragone.etal.2018}, The Three Hundred simulations \citep{Qingyang.etal.2020}, or more recent large cosmological volumes ($\sim$Gpc$^{3}$). These includes the Magneticum set of simulations \citep[e.g.][]{dolag.etal.2025}, which has been also extended to include different cosmologies \citep[e.g.][]{Singh.etal.2020}, the Flamingo suite of simulations \citep{braspenning.etal.2024} that have been carried out by varying feedback and cosmology, and the full-physics Millennium TNG simulation of a large cosmological volume \citep{Pakmor.etal.2023}. Quite interestingly, all such simulations are characterized by a relatively low resolution.

Other studies of populations of galaxy clusters from simulations at higher resolution in fact have been carried out, with different strategies to sample the population of clusters, improve the treatment of AGN feedback, and calibrate the choice of the feedback parameters \citep[see e.g.][for a review]{Valentini.Dolag.2025}. These include a higher resolution version of the {\tt Dianoga} simulations \citep{bassini.etal.2020}, the IllustrisTNG project \citep{nelson.etal.2019, pillepich.etal.2018} that introduced a dual-mode AGN feedback model \citep{Weinberger.etal.2018} that operates in both thermal and kinetic modes, and the TNG-Clusters suite \citep{Nelson.etal.2024}, that has extended IllustrisTNG with a large sample of 351 zoom-in cluster simulations. The FABLE simulations \citep{henden.etal.2018, henden.etal.2020} were specifically calibrated to match the gas fractions and X-ray scaling relations of clusters, utilizing high-entropy bubbles to mimic AGN-driven outflows. Furthermore, the recent PICO-Cluster project included also the effect of magnetic fields, anisotropic heat transport and viscosity \citep{Berlok.etal.2026}. On the high-resolution end, the Hydrangea set \citep{bahe.etal.2017} included zoom-in simulations of 29 massive clusters, and the Romulus-C simulation \citep{Tremmel.etal.2019} provided an extremely detailed zoom-in of a single $10^{14}$~M$_{\odot}$ cluster, featuring a more sophisticated treatment of BH dynamics and the effects of unresolved dynamical friction, which is essential to accurately describe the evolution of SMBHs and of the ensuing AGN feedback.

In fact, such simulations at higher resolution often struggle to simultaneously reproduce the observed galaxy stellar mass function (GSMF) and the thermodynamical profiles of cluster cores. While the Magneticum and Flamingo suites show that current AGN implementations can provide realistic cool cores, the self-regulation between cooling and feedback becomes increasingly difficult to achieve as numerical resolution is increased. Furthermore, several sets of simulations still overpredict the masses of BCGs at the cluster scale, even when they successfully match the properties of smaller groups \citep[e.g.][]{bahe.etal.2017,pillepich.etal.2018,henden.etal.2020}.  

In this paper we aim to address the challenges that cosmological hydrodynamical simulations have in reproducing the feedback-cooling loop in galaxy clusters by combining a relatively high numerical resolution, comparable to that of the TNG-Clusters set, and a critical assessment of numerical effects arising from the coupling of sub-resolution models for SMBH evolution, AGN feedback and star formation from a multi-phase inter-stellar medium. Our analysis is based on the {\tt Dianoga} set of hydrodynamic simulations, consisting of 28 Lagrangian regions which include a wide range of halo masses, with 74 systems having $M_{200} \ge 10^{14}$~M$_{\odot}$ and 23 with $M_{200} \ge 10^{15}$~M$_{\odot}$, at a mass resolution that improves by a factor of 25 with respect to the original set presented in \cite{rasia.etal.2015}. These zoom-in initial conditions are complemented by a relatively small cosmological box, of comoving size of $\simeq 50\,h^{-1}$Mpc, simulated at the same resolution of the zoom-in regions, so as to provide a sort of field reference. These simulations are carried out with the \OG\ code (\citealt{groth.etal.2023, Damiano.etal.2024}, Dolag et al. 2026, in preparation). 

Instead of anchoring our simulations on a number of different observable properties of SMBH, galaxy population and ICM, we rather adopt a minimal strategy by requiring our simulations to reproduce only the relationship between SMBH mass and stellar mass of the host galaxies for the population of field galaxies, while considering other observables as predictions. The origin for any discrepancy between such predictions and observations will be analysed by changing different aspects of the implementation of the purely thermal AGN feedback implemented in our simulations. In total, we explore seven different models in which specific aspects of SMBH accretion and AGN feedback are varied. 

The plan of the paper is as follows. Sect. \ref{sec:simul} describes the simulations analysed in this paper: after presenting the initial conditions in Sect \ref{sec:IC}, we briefly describe the model of star formation and supernova (SN) feedback in Sect. \ref{sec:SF} and of SMBH evolution and AGN feedback in Sect. \ref{sec:AGN}. Sect. \ref{sec:star} presents the main results of our analysis on the stellar content of groups and clusters, by showing the comparison with the observed GSMF (Sect.\ref{sec:gsmf}) and the properties of the BCGs (Sect. \ref{s:bcgs}). Sect. \ref{s:ICM} concentrates on the analysis of the thermodynamical properties of the ICM and the comparison to observational data on scaling relations (Sect. \ref{sec:scal}), and on the radial profiles of such properties (Sect. \ref{s:profs}). The main conclusions of our analysis are presented in Sect. \ref{sec:concl}. An extensive Appendix is dedicated to tracking the origin of anomalous growth of few SMBHs and BCGs (App. \ref{app:OverBH}), to a further comparison between observed and simulated relations between BCG masses and $M_{200}$ (App. \ref{app:bcg}), and to the presentation of results obtained by changing different aspects of the implementation of SMBH accretion and AGN feedback (App. \ref{app:feed}).

\section{Simulations}
\label{sec:simul}
In this section we describe the set of simulations used in this work, which have been carried out with the TreePM-SPH \OG\ code. \OG\ (Dolag et al. 2026, in preparation) represents an evolution of the {\tt GADGET3} code, which in turn provides an improvement of the publicly available {\tt GADGET2} code \citep{Springel2005}. It adopts a mixed MPI/OpenMP parallelization that allows the code to optimally exploit the parallelism offered by computing nodes with a significant amount of shared memory, while reducing the communication surface. Hydrodynamics in \OG\ can be described either by the implementation of Smoothed Particle Hydrodynamics (SPH) described by \cite{beck.etal.2016}, that we adopt for the simulations presented here, or by the Meshless Finite Mass (MFM) scheme described by \cite{groth.etal.2023}. In the SPH formulation by \cite{beck.etal.2016}, the inclusion of a higher-order Wendland-C6 interpolating kernel \citep{DehnenAly12}, of a time-dependent artificial viscosity and of an artificial thermal diffusion \cite{Price2008} alleviate most of the limitations of classical SPH implementations, most notably the capability to treat discontinuities and the development of gas-dynamical instabilities.  

Besides outlining the numerical details of the simulations, we describe here below the model of star formation and AGN feedback, along with the observational constraints used to calibrate the relevant parameters of the AGN feedback.

\subsection{Initial conditions and numerical set-up}
\label{sec:IC}
The {\tt Dianoga} set of simulated galaxy clusters consists of 28 Lagrangian regions, surrounding as many massive halos originally identified in a parent cosmological box having size of $1\,h^{-1}$comoving Gpc (cGpc) \citep{Bonafede11}. This box was originally simulated at low resolution using 1024$^3$ DM particles and assuming a $\Lambda$CDM cosmology with parameters $\Omega_m = 0.24$, $\Omega_b = 0.0375$ for the total matter and baryon density parameters, $h=0.72$ for the Hubble parameter, $n_s =0.96$ for the primordial spectral index, $\sigma_8 =0.8$ for the power spectrum normalization. Zoomed-in initial conditions for such clusters have been then generated using the ZIC code \citep{Tormen96}, which increases mass resolution and adds the relevant high-frequency Fourier modes within a Lagrangian regions surrounding the target halos. Such Lagrangian regions have been chosen so that no low-resolution particles are found by $z=0$ out to at least 5 virial radii from the center of the target cluster. The resulting size of each Lagrangian region is then large enough to include other interesting cluster- and group-sized halos which are not contaminated by low-resolution particles within their virial radius. In the high-resolution region, gas particles are also added in such a way that the ratio of gas and DM particles masses reflects the cosmic baryon fraction. 

The {\tt Dianoga} simulations presented in this paper have a resolution of $m_{\rm DM}\simeq 3.38\times 10^7\msun$ for the mass of DM particles and $m_{\rm gas}\simeq 6.24\times 10^6\msun$ for the initial mass of gas particles. The Plummer-equivalent gravitational softenings are $\epsilon_{\rm DM}=3$ h$^{-1}$ckpc, fixed in physical units below redshift $z=2$ and in comoving units at higher redshift, while $\epsilon_{\rm gas}=1$ h$^{-1}$ckpc at all redshifts. As for the stellar particles, generated by star-forming gas particles, and for BH particles (see below), we adopt $\epsilon_*=250\,h^{-1}$cpc, this more aggressive softening being motivated by the relatively cold dynamics of stellar particles which are generated from the dissipative collapse of gas particles.

Besides these regions, we also simulated a cosmological box having size of $49.26 \,$h$^{-1}$cMpc. Using 576$^3$ DM particles and as many gas particles in the initial conditions, this box size corresponds to the same mass resolution of the {\tt Dianoga} regions. Simulations of this box, which have been carried out assuming the same cosmology and adopting the same star formation and feedback model as the {\tt Dianoga} regions, provide the ''field'' counterpart of the cluster environment, besides including a fair number of halos corresponding to groups and relatively poor clusters.  

We summarize in Table \ref{t:regions} the basic information on the 28 zoomed-in regions and of the cosmological box analysed in this paper\protect\footnote{The ID of the D12 region is missing from this list since initial conditions for this region at the resolution considered here were corrupted and not recoverable.}, including the values of $M_{200}$ and of $R_{200}$\protect\footnote{Here and in the following, $M_{\Delta}$ is defined as the total halo mass contained within the radius $R_\Delta$, encompassing an average density equal to $\Delta \rho_c(z)$, where $\rho_c(z)=3H^2(z)/(8\pi G)$ is the critical density of the Universe at redshift $z$.} of the target halo of each region and of the first two clusters of the Box (Columns 2 and 3). We list also the number of halos with $M_{200}\ge 1.5 \times 10^{13}\mathrm{M_\odot}$ within each region that are free of low-resolution contaminant DM particles. In total, our set of simulations 
includes 293 halos above this mass threshold, out of which 74 and 23 have mass larger than $M_{200}\ge 10^{14} \mathrm{M_\odot}$ and $10^{15} \mathrm{M_\odot}$, respectively.

\begin{table}
\caption{Main properties of the simulations analysed at $z=0$}
\begin{center}
  \begin{tabular}{lrccl}
  \hline \hline
  Region & $M_{200}$ & $R_{200}$ & $N_{13}$ & Models\\
   &  [$10^{14}M_{\odot}$] & [Mpc] &   \\
  \hline
  D1  & 18.61 & 2.49 & 6  & M1,M2,M7 \\ 
  D2  &  4.87 & 1.59 & 6  & M1,M2,M3,M4,M5,\\  
      &       &      &    & M6,M7 \\ 
  D3  &  7.36 & 1.83 & 4  & M1,M2,M7 \\
  D4  &  5.70 & 1.68 & 2  & M1,M2,M4,M5,M6,M7 \\
  D5  &  1.79 & 1.14 & 3  & M1,M2,M7 \\
  D6  & 15.19 & 2.33 & 12 & M1,M2\\
  D7  & 16.58 & 2.40 & 15 & M1,M2\\
  D8  & 18.04 & 2.46 & 14 & M1,M2,M4\\
  D9  &  1.44 & 1.06 & 3  & M1,M2,M3,M4,M5 \\
      &       &      &    & M6,M7 \\ 
  D10 & 16.51 & 2.39 & 10 & M1,M2,M7\\
  D11 & 13.17 & 2.22 & 14 & M1,M2\\
  D13 & 17.89 & 2.46 & 15 & M1,M2\\
  D14 & 20.07 & 2.55 & 7  & M1,M2,M4 \\
  D15 & 19.65 & 2.54 & 11 & M1,M2\\
  D16 & 38.79 & 3.18 & 21 & M1,M2\\
  D17 & 17.90 & 2.46 & 4 & M1,M2\\
  D18 & 12.03 & 2.15 & 12 & M1,M2,M7\\
  D19 & 16.19 & 2.38 & 12 & M1,M2\\
  D20 & 19.83 & 2.54 & 12 & M1,M2\\
  D21 & 17.21 & 2.43 & 5  & M1,M2\\
  D22 & 21.92 & 2.63 & 14 & M1,M2\\
  D23 & 15.39 & 2.34 & 8  & M1,M2\\
  D24 & 14.93 & 2.31 & 4  & M1,M2\\
  D25 & 11.39 & 2.11 & 5  & M1,M2\\
  D26 & 15.97 & 2.37 & 9 & M1,M2\\
  D27 & 18.11 & 2.47 & 4  & M1,M2\\
  D28 & 22.39 & 2.65 & 20 & M1,M2\\
  D29 & 17.17 & 2.42 & 16 & M1,M2\\
Box   & 1.76  & 1.13 & 25 & M1,M2\\
      & 1.69  & 1.12 & -- & --\\
\hline
\end{tabular}
\end{center}
\tablefoot{From
  left to right, the columns report the region ID, the value of $M_{200}$ and of $R_{200}$ for the target cluster of each regions, the number $N_{13}$ of clusters with $M_{200}\ge 1.5 \times 10^{13}$~M$_{\odot}$ within each region not contaminated by low-resolution particles within $R_{200}$ (for the cosmological box we include the information for the two most massive clusters, and include the value of $N_{13}$), the models simulated for each region (see Table \protect\ref{t:models}).}
  \label{t:regions}
\end{table}

\subsection{Star formation and stellar feedback}
\label{sec:SF}
The simulations presented here include a treatment of a metallicity-dependent radiative cooling and heating/cooling from a spatially uniform, redshift-dependent ionizing UV background, following \cite{Wiersma2009}. Star formation and stellar feedback associated to galactic outflows are described according to the model originally presented in \citet[][SH03 hereafter]{Springelhernquist2003}. Within this sub-resolution description of the interstellar medium (ISM), we assume that gas particles denser than $n_H=0.1$ cm$^{-3}$ become multi-phase and star forming. Within each multi-phase particle a cold and a hot gas phase co-exist in pressure equilibrium, with the cold phase providing the reservoir for star formation, that proceeds according to a Schmidt-Kennicutt law \citep[e.g.][]{Kennicutt1998}. Such particles become then eligible to stochastically spawn four generations of collisionless star particles, with probability proportional to their SFR (see SH03 for further details). The resulting mass of stellar particles is $m_*\sim 1.5\times 10^6\msun$, its exact value depending on the mass of the parent gas particle that can vary in our model of stellar evolution and chemical enrichment \citep[e.g.][]{Tornatore2007}.

Galactic outflows driven by Type-II SN explosions are assumed to be launched with a velocity of 350\vel, with a mass-load which is twice the star formation rate. 
We adopt \cite{ChabrierIMF2003} as the stellar initial mass function (IMF) and assume that each Type-II, originating from stars more massive than $8\,\mathrm{M_\odot}$, releases $10^{51}$ ergs. These choices for wind velocity, wind mass-load and mass threshold for Type-II SN correspond to assuming that a fraction of about 0.3 of the energy made available by Type-II SN powers such outflows. Once a star-forming particle is stochastically selected to become a wind particle, according to its own SFR, it is decoupled from hydrodynamics and prevented from undergoing cooling \citep[][]{Maio.etal.2011,Maio.etal.2022}, until its density becomes lower than 0.025 times the density threshold for star formation or it travels for at least 0.25 Gyrs after the decoupling. 
This ensures that wind particles convert their kinetic energy into thermal energy sufficiently far from star forming regions to make the resulting stellar feedback more efficient. Furthermore, preventing outflowing particles to undergo cooling ensures that they are not spuriously assigned to very small time-steps, thus avoiding a substantial and unnecessary slow down of the simulations. 

Chemical enrichment is included following the model originally described in \cite{Tornatore2007}. For the chosen IMF, different stellar populations are assumed to produce Type-II, Type-Ia SN and low/intermediate mass stars in asymptotic giant branch (AGB) phase. More specifically we assume Type-II SN to arise from stars with masses in the range 8--40 $\mathrm{M_\odot}$, while more massive stars directly collapse to black holes (BHs), thus not contributing to chemical enrichment. Such different stellar populations produce metals according to the stellar lifetimes proposed by \cite{PadovaniMatteucci1993}. As for stellar yields, we assume those by \cite{Thielemann2003} for Type-Ia SN, the mass- and metallicity-dependent yields by \cite{Karakas2010} for AGB stars, and the yields by \cite{Nomoto2013} for Type-II SN.  

\subsection{Super-massive BH model and AGN feedback}
\label{sec:AGN}
The AGN feedback model is inspired by the original implementation presented in \cite{Springel.etal.2005} with the later modifications introduced in \cite{Steinborn.etal.2015}.

We list in Table \ref{t:models} the different AGN feedback implementations for which we carried out our simulations. The meanings of the different parameters varied are explained in this section. Considering M1 as a reference model, for all the other models we mark with bold characters the variation(s) with respect to M1. Also, in Table \ref{t:regions} we report for each region the models that have been simulated. We note that both M1 and M2 have been simulated for all the regions. The results presented in the main text will be mainly based on M1, while the effect of implementing the other models will be discussed in the Appendices. 

Intermediate mass BHs are seeded within halos which are identified by executing run-time a friend-of-friend (FoF) group finder on the star particles. A BH is seeded whenever a  stellar FoF group without a BH particle reaches a minimum stellar mass of $\rm 1 \, \times 10^{9} \, h^{-1} \mathrm{M_\odot} $, it has a stellar-to-DM mass ratio $M_*/M_{\rm DM}\ge 0.05$  and a gas-to-stellar mass ratio is $M_{\rm gas}/M_*\ge 0.1$. For zoom-in simulations, we also require that the seeding FoF contains no contaminating low-resolution particles. These conditions guarantee that BHs are seeded in halos where star formation has already taken place and with enough gas to feed the BHs. 

Whenever seeded, a BH is assigned a minimum mass of $2\times 10^5\mathrm{M_\odot}$, scaling linearly with the stellar mass of the seeding halo, and is located at the position of the highest-density gas particle within the FoF group. With such a small mass, a BH particle at seeding is lighter than surrounding particles. As a consequence, two-body scattering with nearby particles and the lack of a proper description of dynamical friction, may cause BH particles to escape the host halo soon after the seeding \citep[e.g.][]{chen.etal.2022}. To overcome this problem, we include in our simulations the model by \cite{Damiano.etal.2024} that accounts for the unresolved dynamical friction exerted by stellar and DM particles (see also \citealt{Damiano.etal.2025}). This sub-resolution model for dynamical friction has been shown to be quite effective in preventing the formation of a spurious population of wandering BHs. 

Once seeded, a BH can increase its mass by gas accretion and by merging with another BH. Gas accretion proceeds according to an Eddington-limited, Bondi-like criterion. As for the Bondi accretion rate, we use the expression
\begin{equation}
    \dot M_B\,=\,\frac{4\alpha \pi G M_{\rm BH}^2 \rho_g}{(c_s^2+v^2)^{3/2}}\,,
\label{eq:bondi}
\end{equation}
where $M_{\rm BH}$ is the BH mass and $\rho_g$, $c_s$ and $v$ are, respectively, the local gas density, sound speed and relative velocity between gas and BH, all estimated at the BH position. As for the $\alpha$ parameter, we follow \cite{Steinborn.etal.2015} and assume that this boost factor takes distinct values to account for the accretion of cold and hot gas \citep[e.g.][]{gaspari.etal.2013}. The value of $\alpha$ determines, to a first approximation, the rapidity with which a seeded BH catches up on the relationship between BH mass and stellar mass of the host galaxy (see below), before AGN feedback starts regulating gas accretion.

\begin{table}
\caption{AGN feedback models}
\centering
\begin{tabular}{lccccc}
  \hline \hline
  Model & $\epsilon_f$ & $(\alpha_h,\alpha_c)$ & Eddington & Evap. & Regions\\
  \hline
    M1  & 0.10 & (10,100) & yes & no & 29 \\
    M2  & {\bf 0.05} & (10,100) & yes & no & 29 \\
    M3  & {\bf 0.05} &  {\bf (3,30)}  & yes & no & 2 \\
    M4  & {\bf 0.05} & (10,100) & yes & {\bf yes} & 5\\
    M5  & 0.10 & (10,100) & yes & {\bf yes} & 3 \\
    M6  & {\bf 0.20} & (10,100) & yes & {\bf yes} & 3 \\
    M7  & {\bf 0.05} & (10,100) & {\bf no} & no & 8 \\
\hline
\end{tabular}
\tablefoot{Column 1: ID of the model; Column 2: BH feedback efficiency $\epsilon_f$ (see Eq. \ref{eq:Eh}); Column 3: values of the cold and hot boost factors for the Bondi accretion (see Eq. \ref{eq:bondi}); Column 4: enforcement of the Eddington limit (see Eq. \ref{eq:edd}) in the BH accretion rate; Column 5: implementation of the evaporation of cold clouds; Column 6: number of regions, as reported in Table \ref{t:regions}, for which each model has been simulated.}
  \label{t:models}
\end{table}
The cold gas accretion is contributed by cold single-phase gas particles with $T<T_c=5\times 10^4 K$ and by the cold component of the multi-phase gas particles. Therefore, the total Bondi accretion rate contributed by hot and cold gas is
\begin{equation}
   \dot M_{\rm B}\,=\,\dot M_{B,h}+\dot M_{B,c}\,,
    \label{eq:bondi_tot}
\end{equation}
where the two terms in r.h.s. of the above expression correspond to the hot and cold accretion rates, as obtainable from Eq. (\ref{eq:bondi}) when using the $\alpha_h$ and $\alpha_c$ values for the boost factor. In the following, we will use $(\alpha_h, \alpha_c)=(10,100)$ in our reference runs (see Table \ref{t:models}). We will also test the effect of assuming instead the lower values $(3,30)$ in the model M3. 

As for the Eddington limit, we assume the expression
\begin{equation}
    \dot M_{\rm Edd}=\frac{4 \pi G M_{\rm BH} m_p}{\epsilon_r \sigma_T c}\,,
    \label{eq:edd}
\end{equation}
where $m_p$ is the proton mass, $\sigma_T$ the Thomson cross-section, and $\epsilon_r$ the radiative efficiency, i.e. the fraction of the rest-mass energy of accreted gas per unit time, which is converted into the radiated luminosity. Therefore, the rate of BH mass change due to accretion is
\begin{equation}
  \dot M_{\rm BH} \,=\,(1-\epsilon_r)\min{(\dot M_{\rm B}, \dot M_{\rm Edd})}\,.
  \label{eq:Mdot}
\end{equation}
In our reference runs, we will assume the Bondi accretion rate of Eq.(\ref{eq:bondi}) to be always Eddington limited. In order to allow for the possibility of super-Eddington accretion \citep[e.g.][]{Lupi.etal.2024,Amorim.etal.2023}, we also carry out a test run in which we removed the Eddington limit (model M7 in Table \ref{t:models}).

As already mentioned, besides by gas accretion, a BH can change its mass by merging with other BHs. Setting the criterion for the BH-BH merging is a well known critical issue in cosmological simulations \citep[e.g.][and references therein]{Wurster2013,Tremmel.etal.2015}. In the simulations presented here, a pair of BHs is instantaneously merged into a single BH whenever the three following conditions are simultaneously met: {\em (i)} the relative velocity between the two BHs does not exceed half of the sound speed computed at the position of the more massive BH; {\em (ii)} the distance between the two BHs does not exceed 5 times the BH gravitational softening; {\em (iii)} the two BHs are gravitationally bound, so that the values of the gravitational potential computed at the position of the two BHs satisfy the condition 
\begin{equation}
    \frac{|\Phi_{\rm BH_1}-\Phi_{BH_2}|}{a} < 0.5 \,c_s^2 -v_{\rm rel}^2\,,
    \label{eq:bound}
\end{equation} 
where $\Phi$ indicates the value of the potential at the position of each BH, $c_s$ is the local sound speed, and $v_{\rm rel}$ is the relative velocity of the two BHs.

A fraction $\epsilon_f$ of the radiated energy, $\epsilon_r\dot M_{\rm BH} c^2$, is used to increase the thermal energy of the surrounding
gas. Therefore, the energy made available in a timestep $\Delta t$ from a BH gas accretion episode
to heat the surrounding gas is
\begin{equation}
  \Delta E_h\,=\,\epsilon_f\epsilon_r\dot M_{\rm BH} c^2\Delta t\,.
  \label{eq:Eh}
\end{equation}
In our simulations, this energy is only used to increase the internal energy of the gas particles surrounding a BH particle, and is distributed using weights given by the SPH interpolating kernel computed at the BH position. In the following, we assume $\epsilon_r=0.1$ for the radiative efficiency. As for the feedback efficiency, we carried out two complete sets of simulations assuming $\epsilon_f=0.05$ and 0.1 (models M1 and M2 in Table \ref{t:models}) in the active accretion quasar mode, i.e. for $\dot M_{\rm BH} > 0.01  \dot M_{\rm Edd}$. In both cases, these values are boosted by a factor of 4 for quiescent accretion with $\dot M_{\rm BH} < 0.01  \dot M_{\rm Edd}$, so as to mimic the more efficient thermal coupling of released energy in the radio-mode phase \citep{churazov.etal.2005,Sijacki.2006,Fabjan.etal.2010}.

Once the energy available from BH accretion is computed, one has to decide how this energy is used to vary the internal energy of surrounding gas particles. While this is straightforward for single-phase gas particles, it is less obvious for multi-phase particles. In the original implementation of the AGN feedback model, multi-phase particles receive their energy and, at the next computation of their density, they decay back on the equation of state of the multi-phase star formation model within a time-scale which is determined by the star formation model (see Eq. 12 of SH03). This leads to an effective loss of the received feedback energy, whenever the updated density still exceeds the star-formation threshold value. Since this happens in the large majority of cases, such a prescription causes a reduction of the AGN feedback efficiency. In order to explore an alternative possibility, we implemented a simple description for the evaporation of the cold component of multi-phase gas particles due to the effect of AGN feedback, that prevents such particles to immediately loose the received energy (model M4 to M6 in Table \ref{t:models}). In Appendix \ref{app:feed} we describe this implementation and show the results of tests that we carried out by implementing it in our simulations.

\begin{figure*}
\centering
\begin{minipage}{.5\textwidth}
\centering
\includegraphics[width=\linewidth]{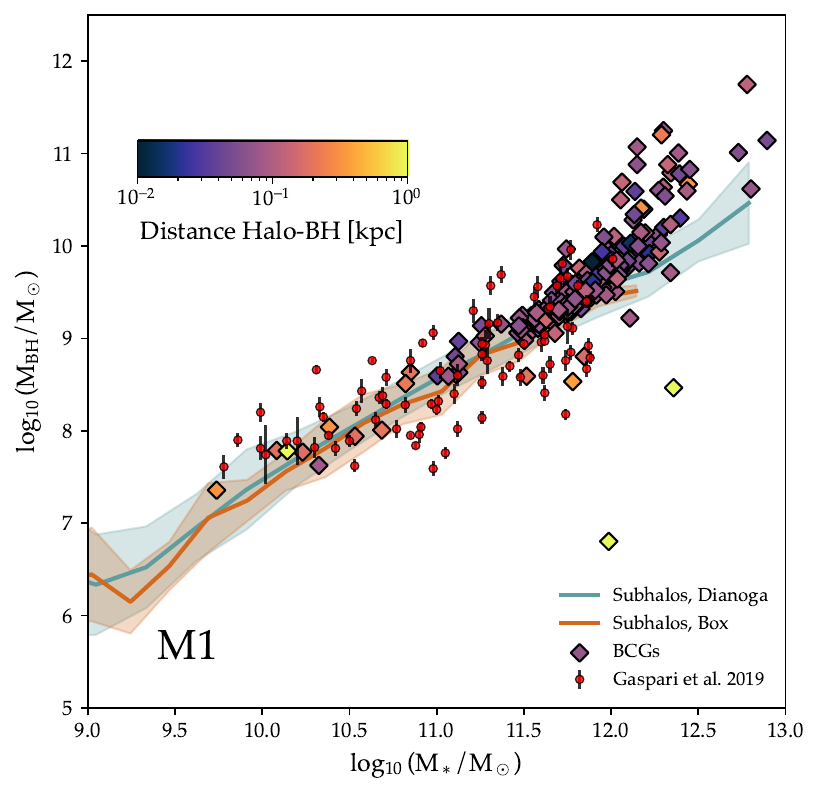}
\end{minipage}%
\begin{minipage}{.5\textwidth}
\centering
\includegraphics[width=\linewidth]{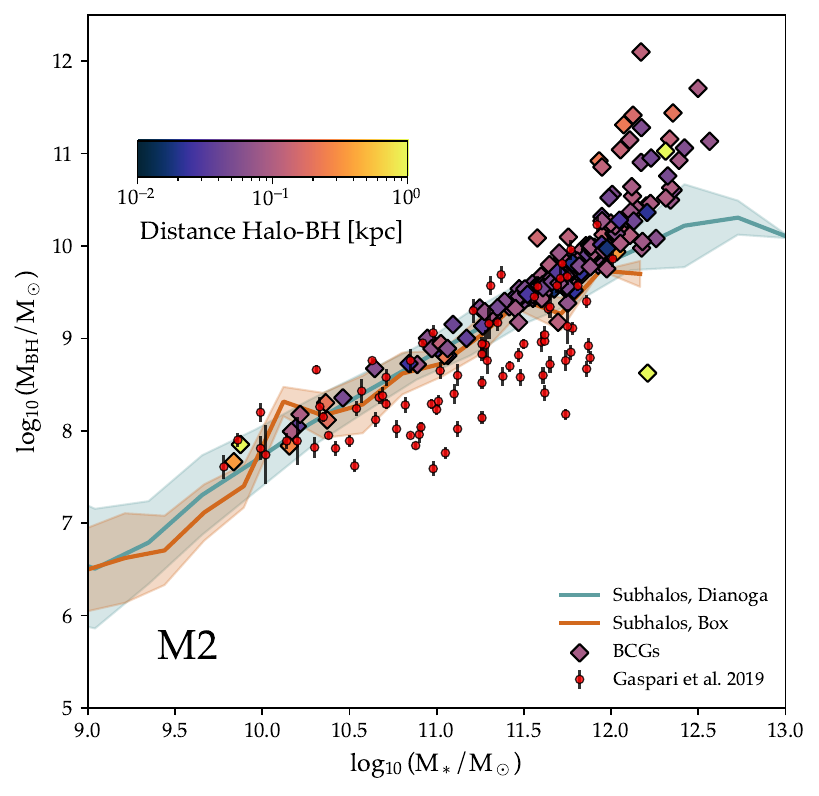}
\end{minipage}
    \caption{Relation between the galaxy stellar masses and the mass of their central SMBHs at $z=0$ for the M1 (left panel) and the M2 model (right panel).  Diamonds refer to central galaxies in simulations and report the most central BH found within 50 kpc from the galaxy center. Stellar masses of central galaxies are computed by summing over all stellar particles within an aperture of 50 kpc. Symbols are color coded according to the distance of the SMBHs from the galaxy center. Shaded areas mark the 16th-84th percentiles for the distribution of the satellite galaxies identified within sub-halos, with the corresponding solid curves indicating the median. Green and red regions correspond to zoom-in {\tt Dianoga} regions and to the cosmological box, respectively. Small red circles with errorbars show observational results for central galaxies from \protect\cite{gaspari.etal.2019}.}
      \label{fig:magorrian}
\end{figure*}

\subsection{Tuning of the model  parameters}
Here and in the following, halos and sub-halos are identified in simulations using \SF\  \citep{Springel.etal.2001} in the version also including gas and stellar particles \citep{Dolag.etal.2009}. In order to select in each of the 28 {\tt Dianoga} regions the BCGs of the target halo and of the secondary halos, we require that the selected halos: {\em (i)} do not host contaminant low-resolution particles within their $R_{100}$; {\em (ii)} have a stellar mass > $10^{11} \, \mathrm{M_\odot}$ within 50 kpc from their center; {\em (iii)} their centers are within a distance of $7\,R_{\rm vir}$ from the target halo; {\em (iv)} have $M_{\rm 200} > 1.5 \times 10^{13} \, \mathrm{M_\odot}$. The most massive halo of each region is flagged as the primary halo, while the others would be the secondary halos. 

In order to associate a BH to a galaxy, for each halo/sub-halo identified by \SF\ we look for the most central BH falling within its half-mass-radius.  For both central and satellite galaxies, stellar masses are computed by accounting for all the stellar particles found within a fixed aperture of 50 physical kpc (pkpc) from the centers identified by \SF. To assess the reliability of the above choices of the parameters regulating gas accretion and BH energy feedback, we show in Fig.\ref{fig:magorrian} a comparison between observational data and simulations on the relationship between stellar mass of galaxies, and the mass of the hosted SMBHs. In Fig.\ref{fig:magorrian}, diamond symbols correspond to all the central galaxies identified in the {\tt Dianoga} zoom-in simulations and in the cosmological box at $z=0$. Symbols are color-coded according to the distance of the BH from the galaxy center. Shaded green area show the 16th-84th percentile in the distribution of sub-halos in these zoom-in simulations, while the shaded red area refers to all galaxies identified in the Box, with the corresponding solid lines marking the median values. As for observations, red circles with errorbars are the observational results for central galaxies from  \cite{gaspari.etal.2019}. As such, they must be compared to results for the simulated central galaxies shown with the diamonds. The left and the right panels refer to the M1 and the M2 model, respectively. By comparing the two, we can appreciate the effect of decreasing the AGN feedback efficiency by a factor of two. Finally, we note very few cases of galaxies hosting a exceedingly low-mass BHs (two cases in M1 and one case in M2). These correspond to few cases in which the massive BH has been temporarily displaced, so that a low-mass BH is assigned as the most centered one, even if it still is at a relatively large distance from the halo center.

We remind here that the tuning of the AGN feedback parameters for the M1 model have been carried out by requiring that results from the Cosmological Box reproduce the observed relationship between BH mass and stellar mass of the host galaxy. In fact, the left panel of Fig.\ref{fig:magorrian} confirms that slope and normalization of the relation predicted by the M1 model agree with observational results, within the scatter. As expected, the lower feedback of the M2 model is less efficient in regulating BH gas accretion, thus producing a higher normalization with respect to data. All the BHs hosted in the central galaxies, with very few exceptions, are well centered, with distances from the galaxy center which is at most comparable to the value of the gravitational softening length. This is obtained thanks to the correction for unresolved dynamical friction, implemented in our simulations, which is quite effective in preventing BHs from wandering away from the center of galaxies \citep{Damiano.etal.2024}.
For both M1 and M2 models, we note no significant difference between central and satellite galaxies and between galaxies in the zoom-in regions and in the Box. On the other hand, we note that the BHs sitting at the center of the most massive BCGs in the zoom-in regions tend to be more massive than expected from the extrapolation of the observed $M_{\rm BH}$--M$_*$ relation, with masses that in a few cases even exceed $10^{11}$~M$_\odot$. As we discuss in Appendix \ref{app:OverBH}, such overmassive BHs are caused by episodes of runaway accretion of high-density cold gas, that our implementation of AGN feedback is not able to preventively remove from the BH surroundings. As expected, decreasing the feedback efficiency in the M2 model makes such episodic runaway accretions more frequent, thus increasing the number of such overmassive BHs. Quite interestingly, at comparable stellar masses, massive galaxies identified as satellites appear to have normal BH masses. We also note in the left panel few cases of exceptionally massive BCGs, with four of them having stellar masses as high as $(5-7)\times 10^{12}$~M$_\odot$. We will discuss such cases in Appendix \ref{app:bcg}. As discussed in Appendix \ref{sec:app_MBHMST}, we verified that BH masses are prevented from reaching exceedingly large values in the models M4-M6, as reported in Table \ref{t:models}, that is when the cold component of the multi-phase gas particles is allowed to evaporate due to AGN feedback (see Figure \ref{fig:mag_comp}).  

We remind that a subset of the {\tt Dianoga} regions simulated for the M2 model have been analysed in previous papers to compare to observational properties of protoclusters: \cite{DiMascolo.etal.2023} for the comparison with the Sunyav-Zeldovich signal detected for the Spiderweb protocluster at $z=2.16$; \cite{esposito.etal.2025} for the study of the galaxy population in protoclusters; \cite{Travascio.etal.2025} for the comparison with the extended X-ray emission associated to a protocluster at $z=3.25$. 

\section{Stellar content of groups and clusters}
\label{sec:star}
In this section we present the results of the analysis of our simulations, and compare them with observational results on the stellar content of clusters and groups, in particular for the GSMF (Sect.\ref{sec:gsmf}) and the properties of the BCGs (Sect \ref{s:bcgs}).

\subsection{The stellar mass function of the galaxy population}
\label{sec:gsmf}
\begin{figure*}
\centering
\begin{minipage}{.5\textwidth}
  \centering
  \includegraphics[width=\linewidth]{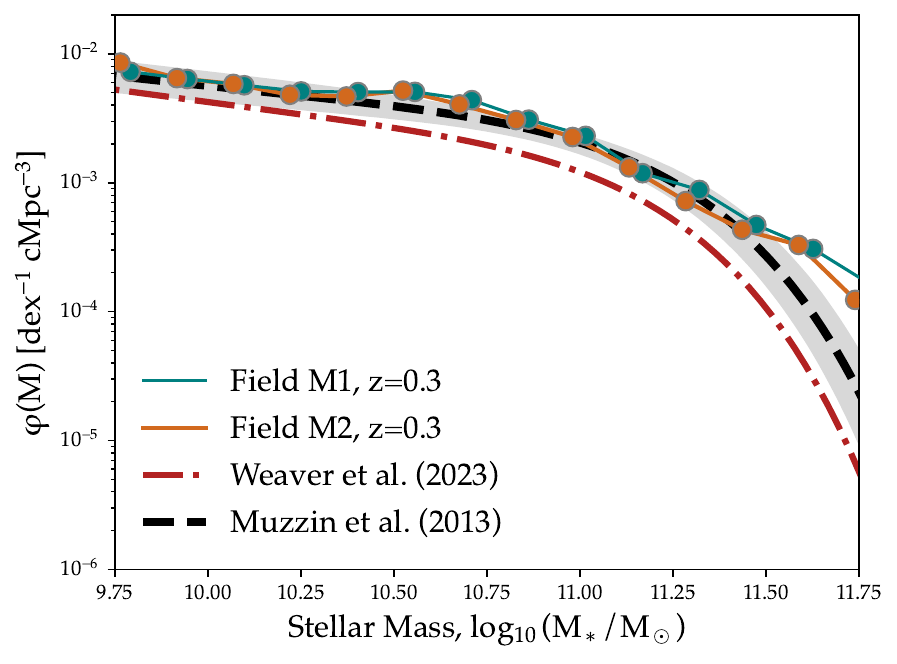}
\end{minipage}%
\begin{minipage}{.5\textwidth}
  \centering
  \includegraphics[width=\linewidth]{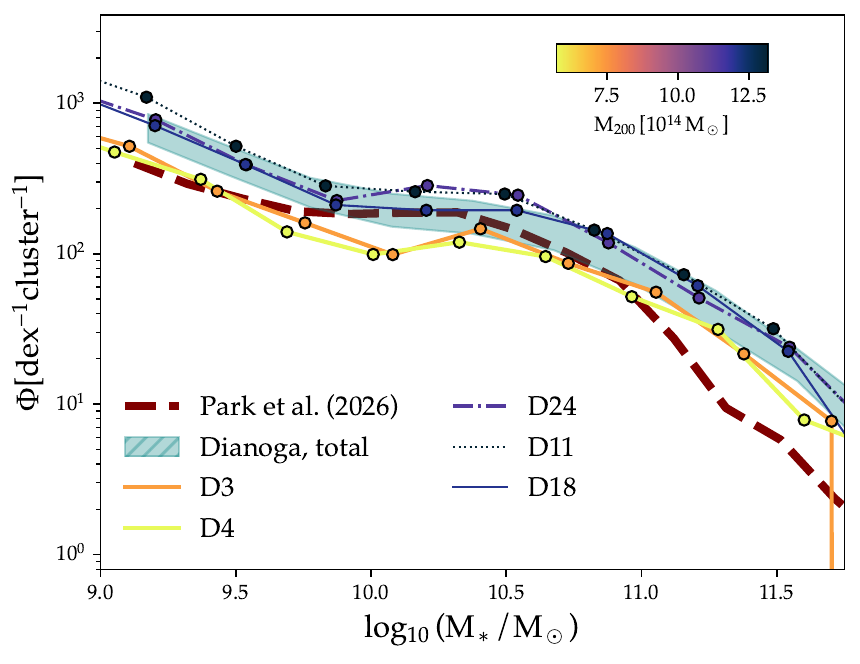}
\end{minipage}
\vspace{-0.5truecm}
\caption{Galaxy stellar mass function (GSMF) in the field (left panel) and in clusters (right panel) for the {\tt Dianoga} simulations and for observations. As for the GSMF in the field, simulation results are shown for both the M1 (green line and filled circles) and M2 (orange line). The GSMF in the field is computed considering all the galaxies in the cosmological boxes. The reported observational results of the field GSMF refer to \protect\cite{Muzzin.etal.2013} (black dashed curve and shaded area) and \protect\cite{Weaver.etal.2023.GSMF} (dash-dotted dark-red curve). As for the field GSMF, the shaded area corresponds to the r.m.s. scatter computed after generating 5000 realizations of the GSMF by sampling the fitting parameters within their confidence intervals, as reported by \protect\cite{Muzzin.etal.2013}. 
As for the GSMF in cluster environment,  we compare results from the {\tt Dianoga} simulations to observational results from \protect\cite{Park.etal.2026} (dashed dark-red line). Curves and circles refer to results for the 5 {\tt Dianoga} regions, whose masses are in the range covered by the observational sample. Circles are color-coded according to the value of $M_{200}$ of the corresponding halo. The light-blue shaded area mark instead the 16-th to 84-th percentile of the whole {\tt Dianoga set}. Stellar mass of galaxies in simulations are computed within an aperture of 50 pkpc.
}
\label{fig:cluster_smf}
\end{figure*}

As a basic observational quantity to assess the capability of our simulations to reproduce a realistic galaxy population, we compare in Fig. \ref{fig:cluster_smf} the predicted GSMF to observational data at low redshift, both for the field (left panel) and within clusters and groups (right panel). As for the GSMF in the field, we show results from the cosmological box simulated for the M1 and M2 models (green curve with circles and orange curve, respectively), both computed at $z=0.3$. Stellar masses of each galaxy identified by \SF\ are computed by assuming an aperture of 50 pkpc. This aperture has been chosen so as to be large enough to encompass at leat five times the observed half-light radius of massive galaxies with $M_\ast \approx 10^{11} {\rm M_\odot}$ \citep[e.g.][]{vanderWel.etal.2014}. We have verified that results are virtually unchanged when using instead a smaller aperture of 30 pkpc.  

Results from our simulations are compared to observational measurement of the low-redshift GSMF from \cite{Muzzin.etal.2013} who analysed the COSMOS/UltraVISTA survey, and from \cite{Weaver.etal.2023.GSMF}, who analysed the COSMOS2020 galaxy catalogue. In this plot results are shown for the range of stellar masses which is covered by observational data. From this comparison, we conclude that our simulations reproduce the shape of the observed GSMF reasonably well at low and intermediate masses, $\log(M_*/{\rm M_\odot})\mincir 11.25$, while they overproduce the number density of higher-mass galaxies. This discrepancy in the high end of the GSMF exists for both the M1 and M2 models. This indicates that the limited efficiency of our AGN feedback model in regulating the masses of the most massive galaxies is not related to the value of $\epsilon_f$, i.e. the parameter formally regulating the AGN feedback efficiency. We note that this scarce efficiency of the implemented AGN feedback should not be merely intended as due to a too low value of the relevant feedback efficience parameters. In fact, we discuss in Appendix \ref{app:feed}, a more efficient feedback can be obtained using the same efficiencies, but allowing this feedback to evaporate multi-phase gas particles (models M4 to M6 in Table \ref{t:models}). 

As for GSMF in the environment of clusters, in the right panel of Fig.\ref{fig:cluster_smf} we compare the predictions from our simulations to the recent observational result by \cite{Park.etal.2026}. In their analysis, they measured the GSMF for nine high-mass clusters in the local Universe using deep Hectospec spectroscopic data from the MACH sample \citep{Sohn.etal.2020}, that allow them to reach spectroscopic completeness down to $\log(M_*/\mathrm{M_\odot})=9.0$. As an interesting results, they found that their GSMF has a steepening at low masses, a feature that is not reproduced by the results from TNG-300 \citep[e.g.][]{pillepich.etal.2018}, which they compare to, while agreeing with such simulation at larger masses, $10.5<\log(M_*/\mathrm{M_\odot})<11.6$. At even larger masses, \cite{Park.etal.2026} found a marginal evidence for an excess of galaxies in TNG-300. Consistently with the results for the field GSMF, we confirm that our {\tt Dianoga} simulations predict a GSMF in such massive clusters that is higher than observed. Quite interestingly, however, our GSMF does predict a steepening in the low mass end, starting at stellar masses comparable to the observed one and with a similar shape. Therefore, while the AGN feedback implemented in our simulations is not efficient enough in regulating star formation in the most massive galaxies, the stellar feedback associated to SN-driven galactic outflows has the required low efficiency to reproduce the large number of low-mass galaxies implied by the results by \cite{Park.etal.2026}.

We point out that a proper comparison between observed and simulated stellar masses would require developing a method to produce mock observations of simulated galaxies, rather than simply measuring them within a fixed aperture. While this goes beyond the purpose of this paper, analyses carried out in this direction  \citep[e.g.][]{Tang.etal.2018,Tang.etal.2021} have shown that including observational effects can in fact have a non-negligible impact and possibly reconcile tensions between observed and simulated GSMF in the high-mass end.

\subsection{Properties of the Brightest Cluster Galaxies}
\label{s:bcgs}
Regulating star formation at the center of the most massive halos is usually considered as one of the main challenges for the models of star formation and energy feedback implemented in cosmological hydrodynamical simulations \citep[e.g.][and references therein]{bahe.etal.2017,pillepich.etal.2018,ragone.etal.2018,henden.etal.2020,Nelson.etal.2024}.  

\begin{figure*}
    \centering
\begin{minipage}{.5\textwidth}
\centering
\includegraphics[width=\linewidth]{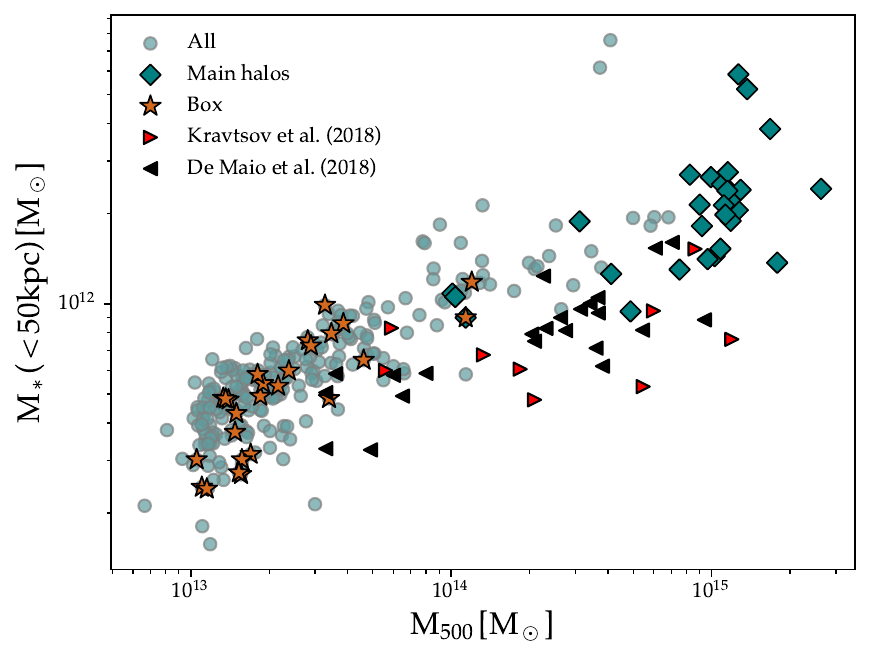}
\end{minipage}%
\begin{minipage}{.5\textwidth}
\centering
\includegraphics[width=\linewidth]{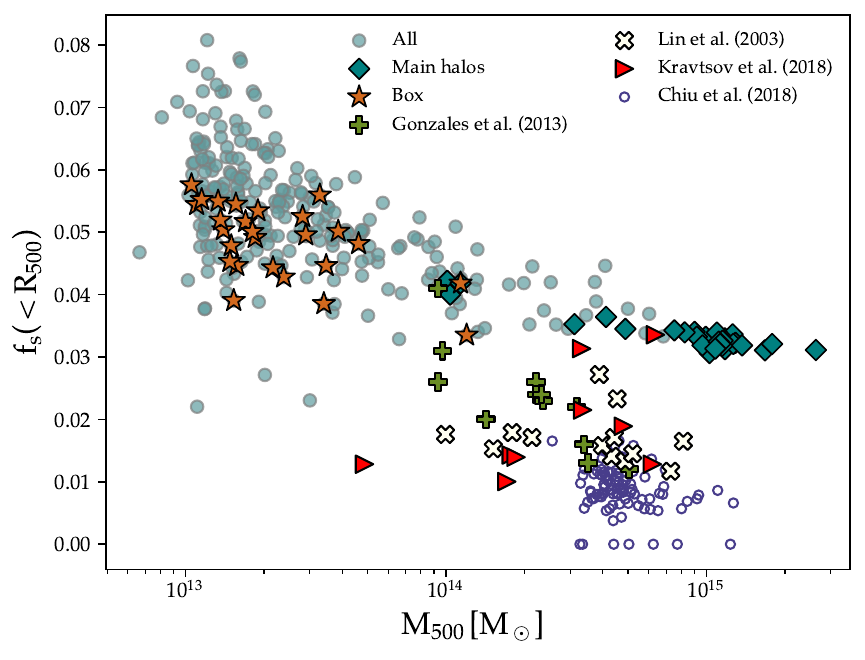}
\end{minipage}
\vspace{-0.3truecm}
    \caption{Stellar mass content as a function of halo mass. Left panel: relation between $M_{500}$ and BCG/BGG stellar mass at $z=0$. As for simulation data, results are shown for the M1 model. Teal diamonds are for the BCGs of the central halos in each {\tt Dianoga} Lagrangian region, light-teal circles are for BCGs/BGGs of secondary halos of such regions, while orange stars are for BCGs/BGGs in the cosmological box. Stellar masses are obtained summing over all stellar particles within a 3D aperture of 50 pkpc. Observational data points are taken from table~4 of \protect\cite{kravtsov.etal.2018} (red triangles) and table~2 of \protect\cite{DeMaio2018} (black triangles). Right panel: stellar mass fraction within $R_{500}$ as a function of $M_{500}$. Symbols for simulation results are the same as in the left panel. As for observational data points, in addition to results from \protect\cite{kravtsov.etal.2018}, we also show those from \protect\cite{lin.etal.2003}, \protect\cite{gonzalez.etal.2013} and \protect\cite{chiu.etal.2018}.}
      \label{fig:m500_mbcg}
\end{figure*}

In the left panel of Fig.\ref{fig:m500_mbcg}  we compare the relationship between  stellar masses of Brightest Group Galaxies (BGGs) and Brightest Cluster Galaxies (BCGs), and $M_{500}$ predicted by our M1 simulations, with the observational results by 
\cite{kravtsov.etal.2018} and \cite{DeMaio2018} (see Appendix \ref{app:bcg} for a comparison to other observational results involving halo masses measured at $R_{200}$). We note that BGGs/BCGs in simulations obey to the same scaling relation, independent of the environment, i.e. of whether they belong to the main halo, to a secondary halo of the zoom-in regions, or whether they are hosted in systems within the cosmological box. Simulation results and observations becomes closer at the scale of groups, $\log (M_{500}/\mathrm{M_\odot})\simeq 13.5$--14, especially when considering the results from \cite{kravtsov.etal.2018}. However, the observed relationship between total halo mass and BCG stellar mass is significantly shallower than in simulations, a results which is shared also by other simulations \citep[e.g.][and references therein]{Biffi.etal.2025}. The large scatter in the observed BCG stellar masses in the richest clusters points toward the presence of significant uncertainties affecting these measurements, which makes difficult to quantitatively assess the level of tension with predictions from simulations.  

These results signal once again the difficulty of our feedback model in regulating star formation in the most massive galaxies. We note that such a limitation is shared by other simulations presented in the literature, and carried out at a comparable resolution \citep[e.g.][]{bahe.etal.2017,pillepich.etal.2018,henden.etal.2020}. In fact, regulating star formation in the most massive galaxies remains one of the challenges of current implementations of galaxy formation models within cosmological hydrodynamical simulations. 

On the other hand, this result is at variance with a previous result presented in \cite{ragone.etal.2018}, which referred to the same set of clusters presented here but simulated at a lower (by a factor of 25) mass resolution. In that analysis, stellar masses of simulated BCGs were found to agree with observational data reasonably well. Quite interestingly the feedback scheme used in our previous analysis is quite similar to that adopted here, with AGN parameters tuned so as to reproduce the relationship between BH masses and stellar masses of the host galaxies. This implies that, in our case, our feedback model responds to the increase in resolution in a way that cannot be simply compensated by suitably changing the values of the relevant parameters, so as to retain the agreement with a reference observable (i.e. the $M_*$-$M_{\rm BH}$ relation in our case). As we discuss in Appendix \ref{sec:app_fst}, lower values of the stellar mass fraction can be obtained by implementing the evaporation of multi-phase gas particles by AGN feedback (models M4 to M6) or removing the Eddington limit in the computation of the BH accretion rates (model M7; see Figure \ref{fig:app_mbcg}). On the other hand, changing the feedback efficiency $\epsilon_f$ or the values of the $\alpha$ boost factors in the expression of the Bondi accretion rate (see Eq. \ref{eq:bondi}) does not lead to any significant change in the BCG stellar masses. The results from these tests suggest that the increase of resolution requires in some cases modifications of the feedback models that go beyond a simple re-tuning of the efficiency parameters. 

\begin{figure}
    \centering
\includegraphics[width=1.\linewidth]{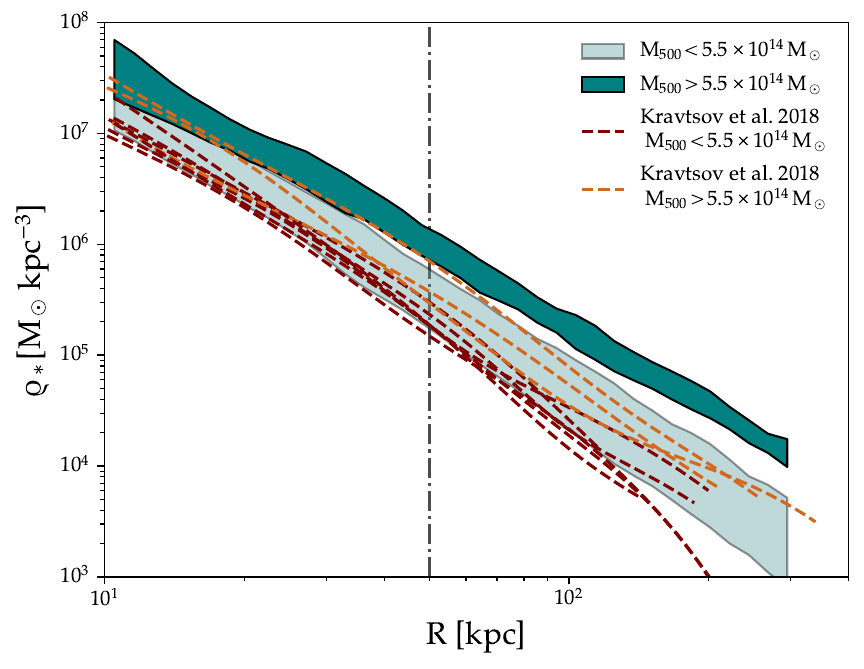}
\vspace{-0.5truecm}
    \caption{Comparison between 3D simulated (shaded areas) and observed (dotted lines) profiles of stellar density. Light green and dark green shaded regions show the $16^{\rm th}$ and $84^{\rm th}$ percentiles of the stellar density profiles for the simulated clusters with mass $M_{500}$ respectively above $5.5\times 10^{14}$M$_{\odot}$ and between $0.5$ and $5.5\times 10^{14}$M$_{\odot}$. Orange and dark-red lines show the stellar density profiles from a set of 8 clusters, measured by \protect\cite[][]{kravtsov.etal.2018}. Each observed profile is plotted out to the outermost radius $R_{\rm out}$, where it has been measured from photometric data. The dot-dashed vertical line indicates the 50 kpc radius which is used to define the BCG boundary. }
    \label{fig:starprof}
    \end{figure}

The question may arise as to whether this excess of the stellar mass of the BCGs is limited to the central part or involves also its outer stellar halo which makes up the diffuse stellar component (DSC). To this purpose, we show in Fig. \ref{fig:starprof} the comparison of our simulation results with the 3D (i.e. deprojected) stellar density profiles from the analysis by \cite{kravtsov.etal.2018}. Stellar density profiles for both simulations and observations  include the contribution from the BCG and its extended stellar envelope, after masking the contribution from satellite (i.e. non-BCG) galaxies. Both observed and simulated clusters are divided into a high-mass bin, with $M_{500}\geq 5.5\times 10^{14}$~M$_\odot$, and a low--mass bin, with $0.5\leq (M_{500}/10^{14}$~M$_\odot) < 5.5$. In general, profiles from simulations are higher than from observations over all the radii covered by observations. Still, observed and simulated profiles have consistent slopes, with a possible evidence for a shallower slope of the observed profiles of high-mass clusters at radii beyond 50 kpc. The agreement of the slopes, despite the different normalizations, confirms that our simulations produce BCGs that are more massive than observed ones, while providing a realistic description of the dynamical processes determining the assembly of the BCG and the formation of the DSC from the disruption of merging galaxies.

In the right panel of Fig. \ref{fig:m500_mbcg} we extend the comparison to the total stellar content of groups and clusters, by showing the total stellar mass fraction within $R_{500}$ as a function of $M_{500}$. In the low-mass end, $M\simeq (1-2)\times 10^{13}{\rm M_\odot}$, we note the presence of groups identified as secondary halos in zoomed regions that are upscattered with respect to groups of similar mass identified in the cosmological volume. This suggests that  effects, possibly related to the timing of halo assembly or interaction with other structures within the overdense environment surrounding massive clusters, could impact star formation at the scale of groups. Since a detailed assessment of this effect would require a dedicated analysis, we defer this to a future analysis. 

In principle, a proper comparison between results from simulations and observational data on the total stellar mass fraction would require reproducing in the analysis of simulations the same assumptions made in data analysis for the components of the total stellar mass budget that are included. Since this is quite difficult to implement, owing to the different assumptions and data used in different analyses, we decided to adopt a conservative approach and include in the analysis of simulations the contribution from all the stars -- BCGs with their DSC  envelopes and satellite galaxies - found within $R_{500}$. As for observational data points, the values reported by \cite{kravtsov.etal.2018} and \cite{gonzalez.etal.2013} include the contribution of BCGs, DSC and galaxies, so that such results can be homogenously compared with our results. In general, this comparison is in line with what indicated in the left panel: simulations tend to produce an exceedingly large amount of stars, by a factor of about 3, in the high-mass end, $M_{500}\sim 10^{15}$~M$_\odot$. Such a discrepancy reduces at the scale of poorer clusters and groups, $M_{500}\lesssim 10^{14}$~M$_\odot$ (see also \citealt{biffi.et.al.2025} for a similar recent analysis based on the Magneticum set of cosmological simulations). 

Quantifying how much of this discrepancy should be ascribed to an inefficient feedback within the massive clusters and how much is contributed by uncertainties in observational data analysis -- e.g. related to SED-fitting assumptions and proper accounting for low-surface brightness DSC -- goes beyond the purpose of this analysis. However, we point out that a tension between predictions from simulations and observed stellar content of the most massive halos is shared by different simulations presented in the literature \citep[e.g.][and references therein]{schaye.etal.2023,Nelson.etal.2024}, thus highlighting the relevance of properly addressing this open problem. Quite interestingly, the tests presented in Appendix \ref{sec:app_fst} show that the models M4-M6, which include the evaporation of the cold phase in star-forming particles, are not only effective in reducing the BCG stellar masses, but also to largely alleviate the tension between {\tt Dianoga} simulations and observations on the total stellar mass fraction in clusters and groups (see Fig. \ref{fig:app_fst}). On the other hand, all the other variants of the reference model, as described in Table \ref{t:models}), do not produce any appreciable variations of the stellar mass fraction.

\section{Properties of the intra-cluster medium}
\label{s:ICM}
In this section we analyse the thermodynamical properties of the ICM for our simulated clusters, and compare them to a selection of observational results. After discussing the results for different scaling relations of total cluster mass against different ICM observables in Sect. \ref{sec:scal}, we analyse the profiles of the ICM thermo-dynamical properties in Sect. \ref{s:profs}. In the following, we will focus on X-ray observable properties of the ICM and thus consider only the hot SPH particles whose internal energy corresponds to a temperature in the range $0.3< T/{\rm keV} < 40$, also excluding all the multi-phase gas particles whose mass fraction in cold gas exceeds 10 per cent.

\subsection{Scaling relations}
\label{sec:scal}
\begin{figure}
    \centering
\includegraphics[width=1.\linewidth]{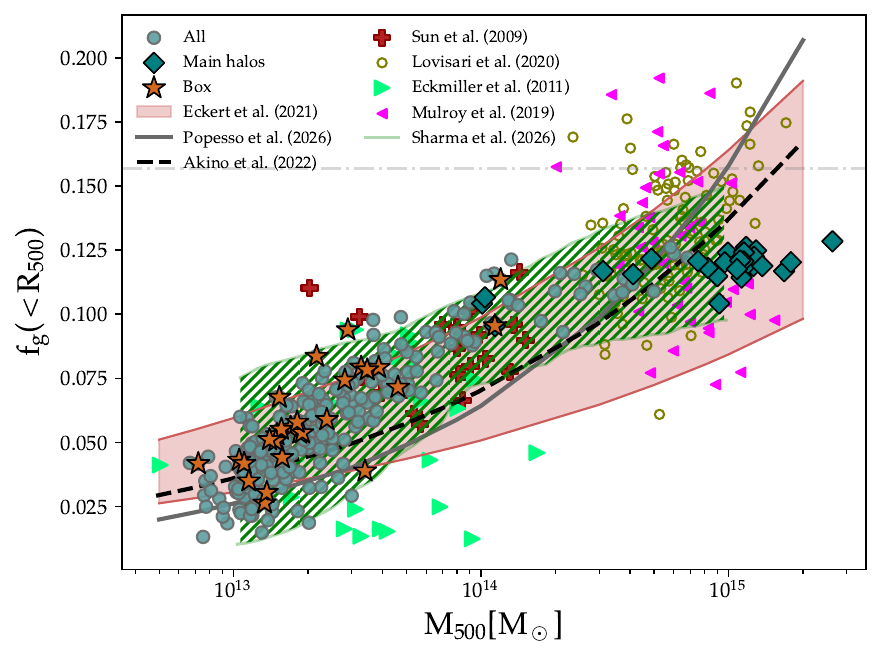}
\vspace{-0.5truecm}
    \caption{Comparison between the total gas content of galaxy clusters and groups predicted by our M1 {\tt Dianoga} simulations and observational results. Results from simulations are shown with the same symbols as in Fig.~\ref{fig:m500_mbcg}. Gas masses from simulated groups and clusters have been obtained by considering only non-star-forming gas particles with temperature $0.3<T<40$ keV. 
    The observational results in the group regime are from \protect\citet[][dark-red crosses]{sun.etal.09} and \protect\citet[][green triangles]{Eckmiller11}, whereas those in the cluster regime are from \protect\citet[][magenta triangles]{mulroy19} and \protect\citet[][yellow circles]{lovisari20}. Best-fit relations to other observational data sets are also shown for \protect\citet[][tick solid line]{Popesso.etal.2026} and \protect\citet[][tick dashed line]{akino.etal.2022}. The green shaded area is from \protect\cite{sharma.etal.2026}, while the shaded red region represents Eq.~(11) of \protect\cite{eckert.etal.2021}.
    All observational data are transformed using our cosmology.
    The grey horizontal dot-dashed line marks the cosmic baryon fraction assumed in our simulations.}
    \label{fig:fgas}
\end{figure}

As a first test, we compare in Fig.\ref{fig:fgas} the simulation predictions on the hot gas fraction of clusters and groups to observational results from different analyses. In fact, the gas content of halos as a function of their mass is recognized as one of the most sensitive diagnostics for the implementation of AGN feedback in cosmological simulations, including its energetics and scale for the energy deposition \citep[e.g.][and references therein]{Bigwood.etal.2025, Davies.etal.2026}. Results for our {\tt Dianoga} simulations are shown for the reference M1 set defined in Table \ref{t:regions}. Teal diamonds are for the central target halos of the 28 {\tt Dianoga} regions, light teal circles for all the other groups and clusters identified in the same regions, while the orange stars refer to groups and clusters identified in the cosmological box. Central clusters have gas fraction which are consistent with those of other non-central objects. At the same time, groups in the box have gas fractions similar to those of groups of comparable mass identified within the {\tt Dianoga} regions, with a marginal tendency for the former to have slightly larger $f_{\rm g}$ values. This suggests that hot atmosphere that characterizes the surroundings of massive clusters has at most a marginal impact on the gas content of groups located in this environment. 

As for observations, we show results from different sources. \cite{sun.etal.09} presented an analysis of 43 nearby groups and of 14 clusters from a sample by \cite{vikhlinin.etal.2009}, all observed in the X--rays with the Chandra satellite (red crosses). \cite{Eckmiller11} combined data for 26 groups with those of the HIFLUGCS clusters from \cite{reiprich.etal.2002} (green triangles). 
\cite{Lovisari15} analysed a set of 82 groups and clusters originally identified in the ROSAT All-Sky Survey and followed-up with XMM-Newton (yellow circles). \cite{mulroy19} analysed a set of 41 clusters at $0.15<z<0.3$, all having weak-lensing HSC mass measurements and gas masses from {\em Chandra} and {\em XMM-Newton} X-ray observations (magenta triangles). \cite{Popesso.etal.2026} carried out a stacking analysis of eFEDS eROSITA data on a sample of galaxy groups and clusters identified in the GAMA survey \citep{robotham.etal.2011} (solid gray curve). \cite{akino.etal.2022} analysed 136 X--ray observations from the scale of groups to massive clusters in the redshift range $0\lesssim z \lesssim 1$, using total masses from HSC weak-lensing  measurements and gas masses from XMM-Newton data (dashed black curve). We show with the shaded red area, the relationship provided by \cite{eckert.etal.2021}, which encompasses a compilation of different $f_{\rm g}$ observational measurements. Finally the shaded green area refers to the recent measurement of $f_{\rm g}$ from \cite{sharma.etal.2026}, who used a compilation of 109 Fast Radio Bursts  with measured redshifts and dispersion measures. 

In general, results from our simulations fall within the scatter of observational data for the most massive clusters. We note that simulations predict a trend of $f_{\rm g}$ with total mass that is rather flat for $M_{500}\magcir 2\times 10^{14}$~M$_\odot$ \citep{rasia.etal.2025}, and with a scatter which is much smaller than that of observational data. Owing to such a large observational scatter, it is not obvious whether the increasing $f_{\rm g}$ at large masses implied by the best-fitting relations by \cite{Popesso.etal.2026} and \cite{akino.etal.2022} is significantly discrepant from the results of our simulations. 

For poorer systems, $\log (M_{500}/\mathrm{M_\odot})\mincir 13.5$, results from our simulations reproduce quite well the observed gas content. Quite interestingly, the relatively low $f_{\rm g}$ values at the scales of groups, $M_{500}\sim 10^{13}\mathrm{M_\odot}$, found by \cite{Popesso.etal.2026} is also in line with recent results based on measurements of the kinetic Sunyaev-Zel'dovich (kSZ) effect from the Atacama Cosmology Survey (ACT) data stacked on DESI galaxies \citep{hadzhiyska.etal.2025,Chaussidon.etal.2026}. In fact, these authors found that groups of $\sim 10^{13}\mathrm{M_\odot}$ at an effective redshift $z\simeq 0.7$ contain a gas fraction within the virial radius as low as $f_{\rm g}/(\Omega_{\rm b}/\Omega_{\rm m})\simeq 0.3$, corresponding to $f_{\rm g}\simeq 0.05$ for our choice of cosmological parameters. Quite interestingly, our results for groups also agree with the observed stellar mass of BGGs (see left panel of Fig. \ref{fig:m500_mbcg}). In general, these results suggest that our model of star formation and feedback produce realistic results at the scale of galaxy groups, without invoking any particularly strong feedback. 

On the other hand, at intermediate mass scales, $M_{500}\sim 10^{14}$~M$_\odot$, our simulations predicts values of $f_{\rm g}$ that lie on the upper end of the range indicated by observations. Overall, the comparison between the gas mass and stellar mass fractions from the {\tt Dianoga} simulations and observations suggests that our feedback model has the required efficiency at the scale of groups, while its efficiency may be too low in more massive systems. Such a weak feedback manifests itself into a marginal overestimate of the gas content of intermediate-mass halos, and in exceedingly larger stellar content - and overmassive BCGs - for the most massive clusters.  

\begin{figure}
    \centering
    \includegraphics[width=\linewidth]{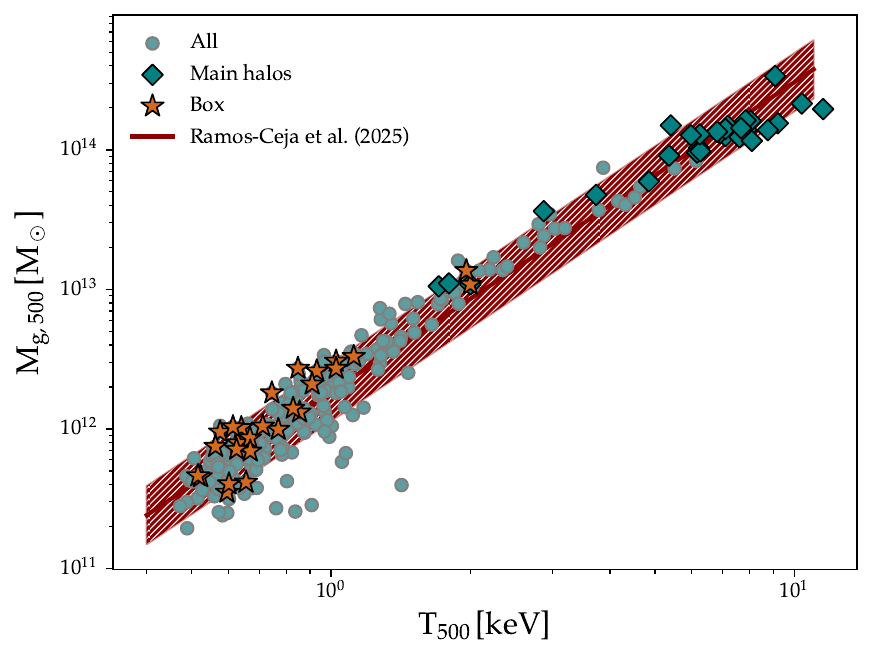}
    \vspace{-0.5truecm}
    \caption{Scaling relation between ICM/IGM gas mass and temperature, both computed within $R_{500}$. Colours and symbols for the {\tt Dianoga} simulations have the same meaning as in the left panel of Fig.\protect\ref{fig:m500_mbcg}. The solid line and the shaded area show the best-fitting relation to observational results for groups and clusters in the eROSITA All Sky Survey from \protect\cite{Ramos.etal.2025}, and the corresponding scatter.}
    \label{fig:mgtsl}
\end{figure}

Fig.\ref{fig:mgtsl} shows the scaling relation between gas mass within $R_{500}$ and the temperature for our {\tt Dianoga} simulations. As for  observational results, we report in this plots the best-fitting relation presented by \cite{Ramos.etal.2025} for about 3000 galaxy groups and clusters selected from the first eROSITA All Sky Survey (eRASS) within the redshift range $0.05<z<1.07$ and mass range $1.1\times 10^{13}<M_{500}/M_{\odot}<1.6\times 10^{15}$ (solid red line and shaded area). In this paper, they parametrized the redshift evolution of the $M_{\rm g}$--$T$ relation with the following expression: $E(z)/E(z_{\rm piv})^{C}$, where $z_{\rm piv}=0.25$. Since their best-fit value for $C$ was found equal to $-1.00$, to compare with their data we multiply our simulated values for $E(z=0.25)$. 

\begin{figure}
    \centering
    \includegraphics[width=\linewidth]{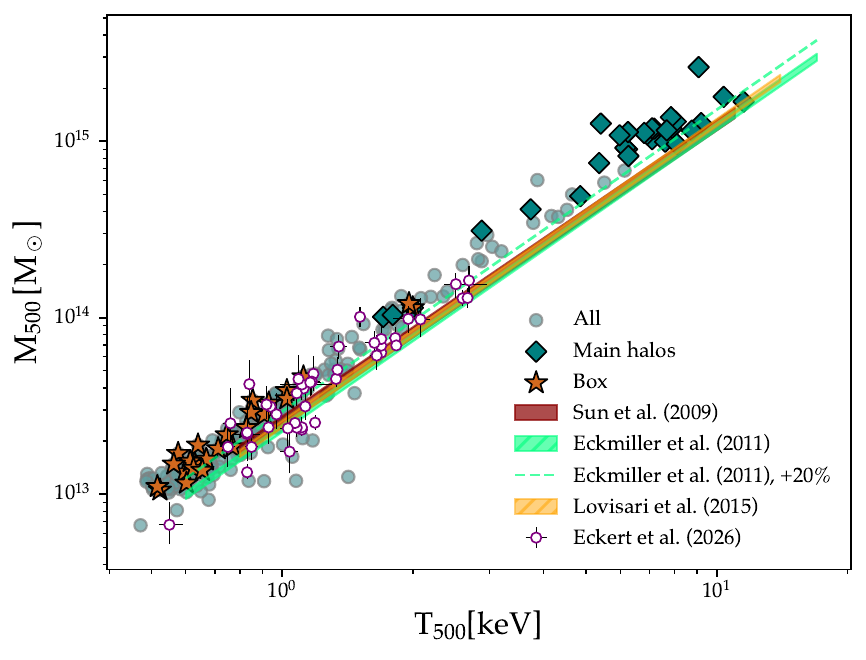}
    \vspace{-0.5truecm}
    \caption{Scaling relation between ICM/IGM temperature and total mass, both computed within $R_{500}$. Symbols and colours for the {\tt Dianoga} simulations are the same as in Fig.\protect\ref{fig:m500_mbcg}. In simulations, the reported temperature corresponds to the spectroscopic-like definition by \protect\cite{mazzotta.etal.2004}. As for observational data, we show results from \protect\citet[][small magenta circles with errorbars]{Eckert.etal.2026}, \protect\citet[][dark-red line]{sun.etal.09}, \protect\citet[][orange area]{Lovisari15} and \protect\citet[][light-green area]{Eckmiller11}. For the latter, we also show with the dotted-dashed green line their best-fitting relation with a normalisation increased by 20 per cent, to show the effect of a possible mass underestimate due to an X-ray mass bias.}
    \label{fig:mtsl}
\end{figure}

Temperature for simulated clusters and groups are computed using the spectroscopic-like formula introduced in \cite{mazzotta.etal.2004}, which has been shown to provide a rather accurate estimate of the X--ray spectroscopic temperature (see also \citealt{vikhlinin.etal.2006}). 
Given that we exclude particles with internal energy below 0.3 keV, 
temperatures of systems with $T_{500}\mincir 1$ keV, reported in Figs. \ref{fig:mgtsl} and \ref{fig:mtsl}, should be taken with some caution, since those systems have a temperature comparable to those of excluded particles.

While this figure conveys in principle the same information as Fig. \ref{fig:fgas}, it shows the relation between two quantities which are both directly observable from X-ray data. Overall, this comparison is in line with that shown in Fig. \ref{fig:fgas}: halo gas masses in the {\tt Dianoga} simulations agree -- with the aforementioned caveat for $T_{500} < 1\,$keV -- with observational results at the scale of groups, they are on the upper end of the observational scatter at $T\sim 3\,$keV, while being on the lower end of this scatter for the hottest systems with $T\simeq 8$--10 keV. While the scaling $M_{\rm g}$--$T$ relation from {\tt Dianoga} shows some curvature with respect to a pure power-law scaling, it is difficult to say whether a similar curvature is hidden in the large large scatter in the observational data by \cite{Ramos.etal.2025} (see their Fig. 9). A dedicated analysis of observational measurement of $M_{\rm g}$ would be required to properly account also for the effect of this scatter. In fact, a proper assessment of whether a curvature or a broken power law for this scaling relation provides a better representation of the data would have interesting implications to understand how feedback acts as a function of halo mass.

As for the scaling relation between total halo mass, $M_{500}$, and X--ray temperature within $R_{500}$ we show in Fig.\ref{fig:mtsl} the comparison between the {\tt Dianoga} predictions and different observational determinations. Also in this case, temperature of simulated clusters and groups have been computed according to the spectroscopic--like definition. At fixed mass {\tt Dianoga} clusters tend to be somewhat cooler than the observed ones. Only at the scale of groups, $T\sim 1\,$keV, our simulations better agrees with the results from \cite{Eckert.etal.2026}, who analyzed high-quality XMM-Newton X--ray data for 44 galaxy groups.  

We note that the masses entering in the observational relations shown in Fig. \ref{fig:mtsl} are all based on the hydrostatic equilibrium applied to X--ray temperature measurements. A number of independent analyses of cosmological hydrodynamical simulations consistently indicate that X--ray masses could be biased low by about 15-20 per cent \citep[e.g.][and references therein]{pratt.etal.2019}. This bias is contributed for a major part by the violation of the assumption of hydrostatic equilibrium \citep[][]{biffi.etal.2016}, with a contribution of X--ray temperature bias \citep{rasia.etal.2006}, given that temperatures measured from X--ray spectral fitting tend to slightly underestimate mass--weighted temperatures \citep{seppi.etal.2026}. To account for the presence of such an X--ray mass bias, we also show in Fig. \ref{fig:mtsl} the result of shifting upward the best-fitting relation by \cite{Eckmiller11} by 20 per cent (dashed green line). While it is beyond the scope of this paper carrying out a detailed analysis to recover hydrostatic masses from X--ray spectroscopic temperatures for our simulated clusters, it is clear that the level of disagreement between the simulated and the observed $M_{500}$--$T_{500}$ relation is comparable with the presence of such a bias in X--ray mass estimates. In principle, correcting upward $M_{500}$ by 20 per cent would imply also revising upward by about 10 per cent the value of $R_{500}$, i.e. the radius within which the temperature is measured. However, owing to the mild temperature gradients in the ICM outskirts, both in simulations and observations (see Sect. \ref{s:profs} below), the correction on $T_{500}$ should be at most marginal. On the other hand, increasing $R_{500}$ has a sizeable impact also on the resulting value of gas mass within this radius. In fact, we verified that the overall effect on the normalization of the $M_{g,500}$--$M_{500}$ relation shown in Fig. \ref{fig:mgtsl} is negligible. For this reason we show the effect of hydrostatic mass bias only for the scaling of temperature with total mass. 

\begin{figure}
    \centering
    \includegraphics[width=\linewidth]{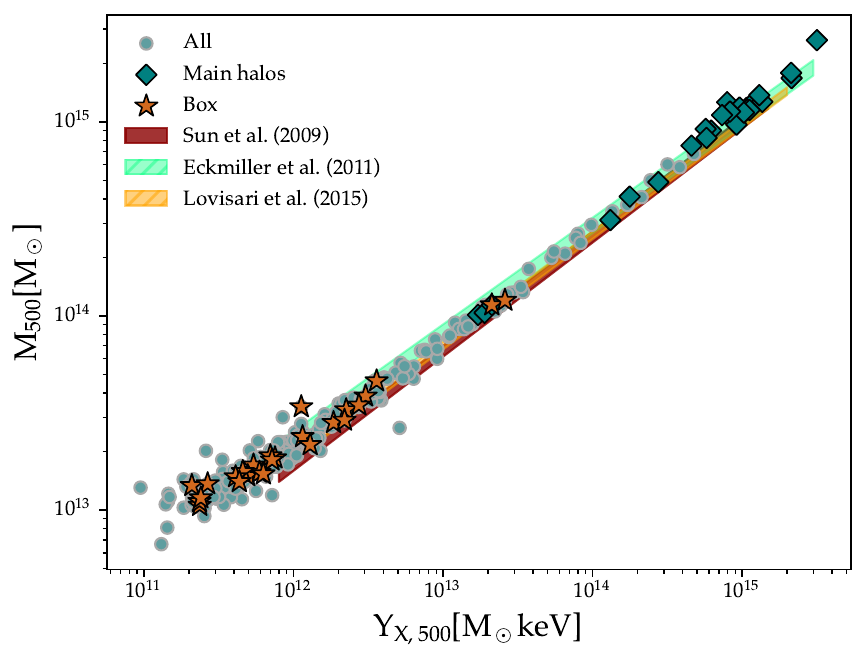}
    \vspace{-0.5truecm}
    \caption{The $Y_{X}$--$M_{500}$ scaling relation for the {\tt Dianoga} clusters and from different observational determinations. Symbols for the simulations and line types for observational results have the same meaning as in Fig. \protect\ref{fig:mtsl}. The values of $Y_X$ from simulations are computed by multiplying gas mass and spectroscopic-like temperature, both computed within $R_{500}$, while excluding the core region inside $0.15\,R_{500}$ for the computation of the latter quantity (see text).
    }
    \label{fig:myx}
\end{figure}

Finally, we show in Fig.\ref{fig:myx} the results on the scaling relation between $M_{500}$ and $Y_{X,500}$. The latter quantity, originally introduced by \cite{kravtsov.etal.2006}, is the X--ray analogue of the Compton-y parameter measured by the Sunyaev-Zel'dovich (SZ) effect, and it is defined as
\begin{equation}
    Y_{X,500}\,=\,M_{\rm g,500}\,T_{\rm ce,500}\,.
    \label{eq:yx}
\end{equation}
Here $T_{\rm ce,500}$ is the core-excised spectroscopic-like temperature within $R_{500}$, computed by removing the contribution from all gas particles within $0.15\,R_{500}$. This observable has been suggested to be a precise and robust cluster mass proxy: it is precise, since its scatter against cluster mass is small, $\mincir 10$ per cent; it is robust since the slope and normalization of its scaling against cluster mass is quite close to the prediction of the self-similar model \citep[e.g.,][]{kravtsov_borgani}, and weakly dependent on the models of star formation and feedback included in the simulations \citep[e.g.,][]{planelles.etal.2017}.

In general, results from simulations are in good agreement with observational results. Quite interestingly, tensions between observations and simulations on the temperature and gas mass scaling relations combine in such a way to produce an excellent agreement on $Y_X$. This confirms the robustness of $Y_X$ as a mass proxy, almost independently of the uncertainties in the detailed modelling of the physical processes regulating the ICM thermodynamics. 

In summary, the results presented in this section highlight that our {\tt Dianoga} simulations predict scaling relations between global ICM/IGM observational properties and total mass which agree quite well with observations. Quite interestingly, these results hold independently of whether groups and clusters are identified in a cosmological box or are identified in the environment surrounding massive clusters. While in general our simulations prove successful in predicting the gas content at the scale of groups and the global thermal ICM content, as measured by the $Y_X$ quantity, they tend to predict a marginal excess of gas content in relatively poor clusters with $M_{500}\sim 10^{14}$~M$_\odot$, and a temperature, at fixed cluster mass, which is slightly lower than observed at the scale of clusters.

\subsection{Profiles of thermodynamical properties}
\label{s:profs}
Besides the scaling relations discussed in the previous section, the profiles of the thermodynamical ICM properties represent the fingerprint of the dynamics of the hierarchical growth of clusters and groups, and of the history of energy feedback from stars and AGN \citep[e.g.,][and references therein]{kravtsov_borgani,gaspari.etal.2019,voit.etal.2023}. 

In this section, we show the comparison between profiles of ICM observable quantities -- electron number density, temperature, entropy and pressure -- from our {\tt Dianoga} simulations and observations. For the purpose of this comparison, we include in the sample of simulated clusters only those identified at $z=0$ and having mass $M_{500}> 10^{14}$M$_{\odot}$. With this selection, we compare them to observational results from the X-COP project by \cite{ghirardini.etal.2019}. The X-COP project \citep{eckert.etal.2019} is based on XMM-Newton observations for a set of 12 nearby galaxy clusters at low redshift, $0.04\mincir z \mincir 0.1$ and masses in the range $3\times10^{14} < M_{500}/{\rm M_\odot} < 1.2\times 10^{15}$, which were already identified in the Planck SZ survey of clusters \citep{Planck_XXIX}. The general approach of this project was to combine X--ray and SZ information with the purpose of mapping profiles of ICM observables, and to recover the hydrostatic mass profiles, from the core regions out to the cluster virial radius \citep[see also][]{ameglio.etal.2009}. Therefore, while based on a relatively limited number of objects, the X-COP data provide a unique mapping of the ICM properties from the central cluster regions, which are mostly sensitive to the effect of AGN feedback, out to the cluster outskirts, where such properties should mostly be driven by gravity and, as such, follow a self-similar scaling.

Following \cite{ghirardini.etal.2019}, we normalise the temperature, entropy\footnote{As usual in the study of the ICM thermodynamics, entropy is defined as $K=T/n_e^{2/3}$ \citep[e.g.,][]{voit2005,borgani_kravtsov2011}, with $T$ being the ICM temperature and $n_e$ the corresponding electron number density.}, and pressure profiles by the following factors:
\begin{equation}
\label{eq:t500}
    T_{500}\,=\,8.85\ {\rm keV} \left(\frac{M_{500}}{10^{15} h_{70}^{-1}M_{\odot}}\right)^{2/3} \left(\frac{\mu}{0.6} \right)\,;
    \end{equation}
    \begin{equation}
P_{500}= 3.426 \times 10^{-3} {\rm keV \,cm^{-3}} \left(\frac{M_{500}}{10^{15}h_{70}^{-1}M_{\odot}} \right)^{2/3} \left(\frac{f_b}{0.16}\right) \left(\frac{\mu}{0.6}\right)\,;      
    \end{equation}
    \begin{equation}
    K_{500}= 1667\ {\rm keV \, cm^{2}} \left(\frac{M_{500}}{10^{15}h_{70}^{-1}M_{\odot}}\right)^{2/3} \left( \frac{f_b}{0.16}\right)^{-2/3} \left(\frac{\mu}{0.6}\right)\,.
\end{equation}

In the analysis of the simulations we assume $\mu=0.588$ as a global value of the mean molecular weight and $f_b=\Omega_b/\Omega_m=0.156$ for the value of the baryon fraction. In the above expressions, we do not include the dependence on $E(z)$, i.e. the factor determining the redshift-dependence of the Hubble expansion rate, since the results of our simulations are considered at $z=0$. The gas density profiles are computed as the ratio between the sum of the gas particle masses and the volume of the spherical shells. As for the temperatures, we use the spectroscopic-like formulation, while for the pressure we apply the SPH correction as discussed in \cite{planelles.etal.2017}. Finally, the entropy profiles are obtained from the median entropy values within each radial bin.

\begin{figure*}
    \centering
    \includegraphics[width=0.45\textwidth]{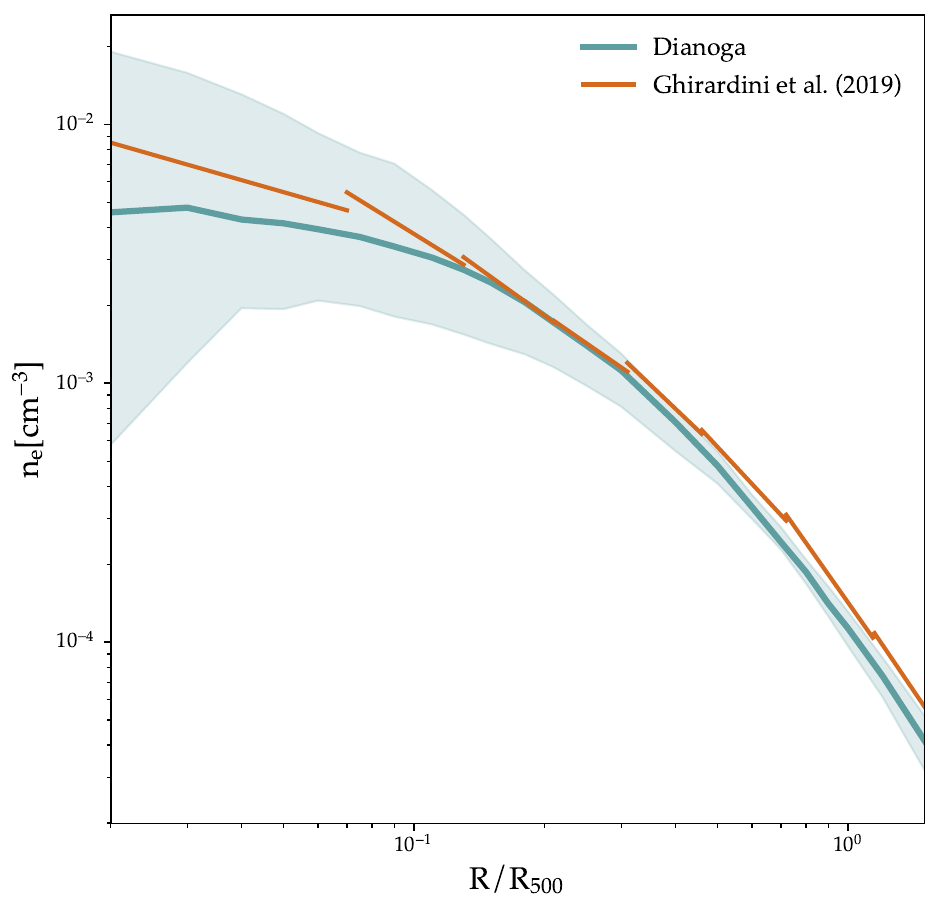}
    \includegraphics[width=0.443\textwidth]{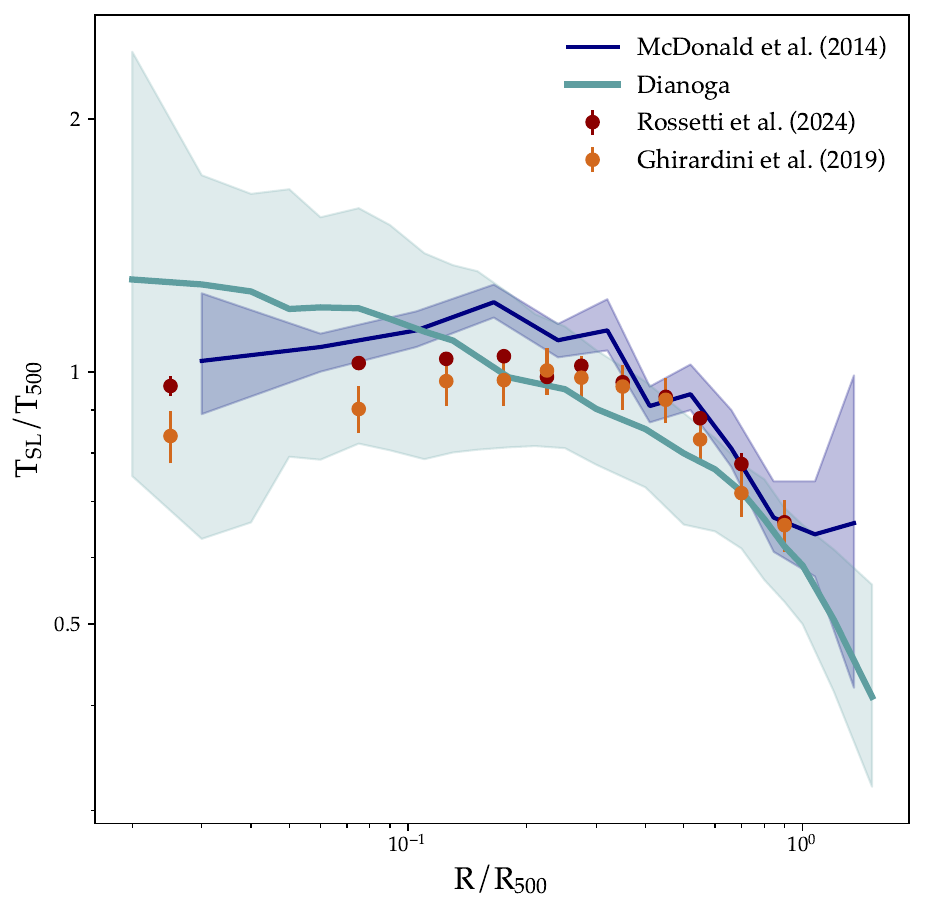}
    \includegraphics[width=0.45\textwidth]{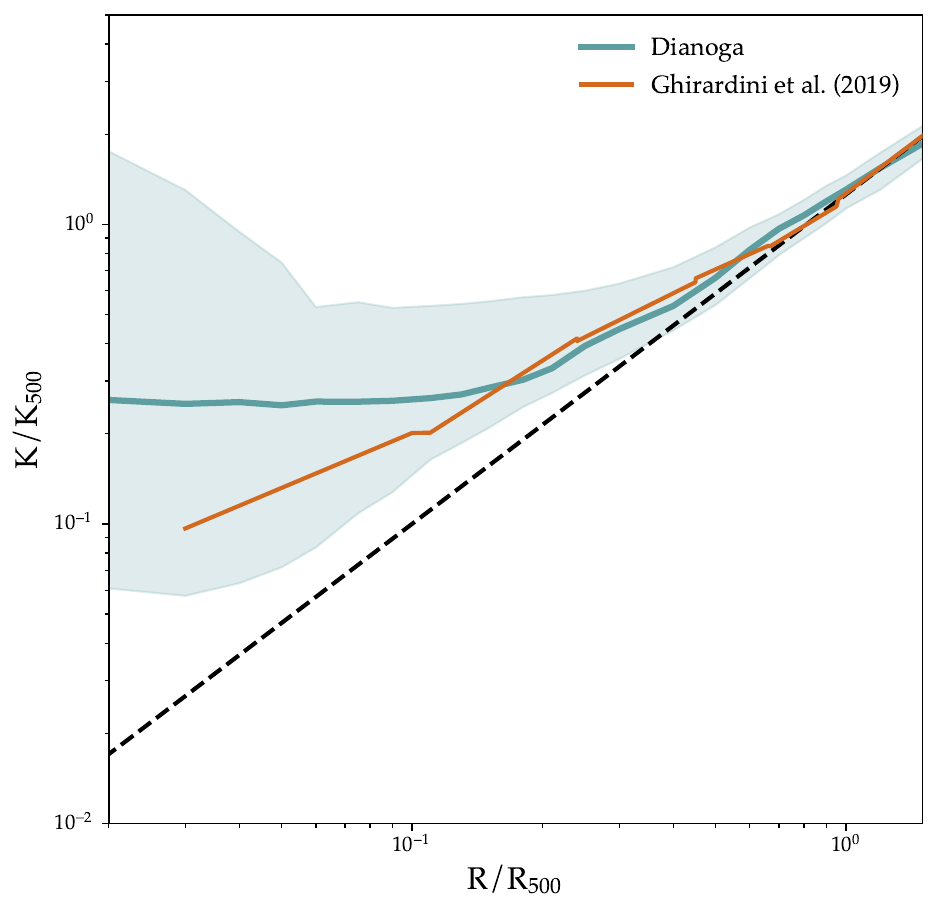}
    \includegraphics[width=0.45\textwidth]{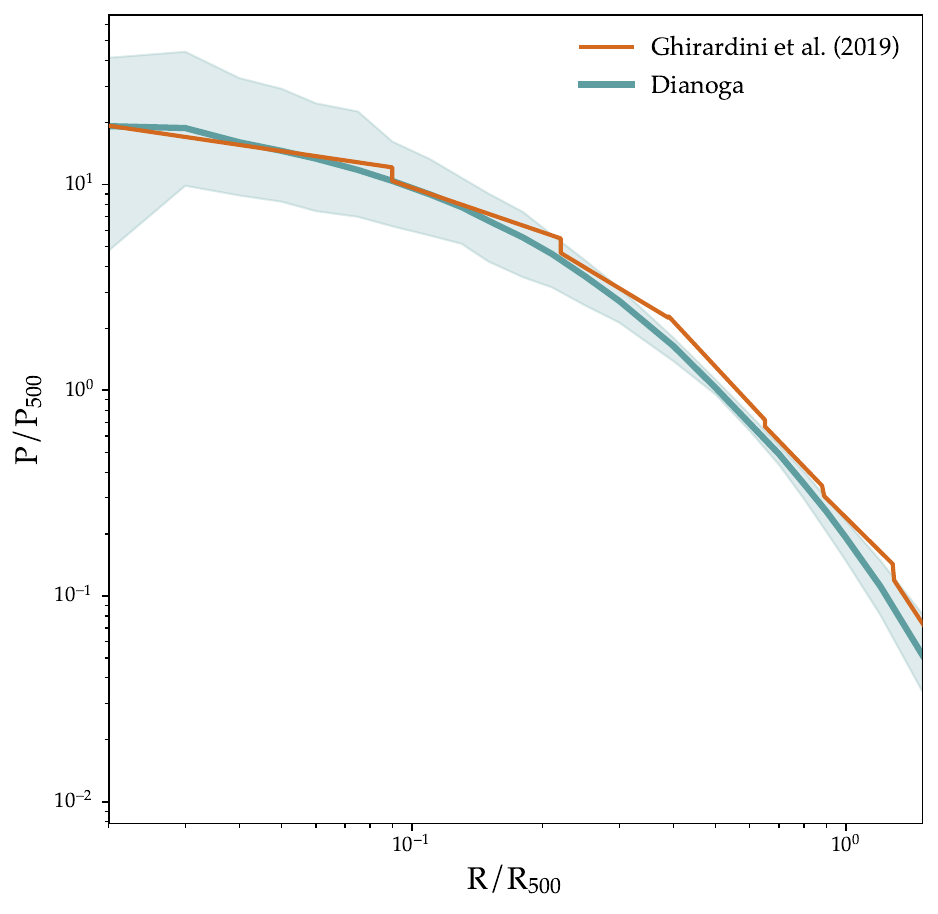}
    \caption{Comparison between electron number density $n_e$ (top-left), spectroscopic-like temperature $T_{\rm SL}$ (top-right), entropy $K$ (bottom-left), and pressure $P$ (bottom-right) profiles from the {\tt Dianoga} clusters (green solid line and shaded areas) and observational data. Profiles are normalized to the values of the corresponding quantity at $R_{500}$, with the exception of $n_e$ that is plotted in physical units. The orange broken curves show the piece-wise power-law fits to the profiles of the observational samples of X-COP clusters, as reported in Table~2 of \protect\cite{ghirardini.etal.2019}. As for simulations, we show profiles of the {\tt Dianoga} clusters with $M_{500}>3 \times 10^{14}$M$_{\odot}$. The green shaded areas include the distribution between 16$^{\rm th}$ and 84$^{\rm th}$ percentiles around the median values of the simulated profiles. Only for the temperature profiles, we also show the results from \protect\cite{mcdonald.etal.2014}, and \protect\cite{rossetti.etal.2024}, along with the data points by \protect\cite{ghirardini.etal.2019}. In the entropy profile panel, the black dashed line show the self-similar relation $K(R) \propto R^{1.1}$ \protect\citep{Tozzi2001,voit.etal.2005}.}
    \label{fig:prof}
\end{figure*}

The results of this comparison are shown in the four panels of Fig. \ref{fig:prof}. In all panels, the green solid curved and shaded area correspond to the median {\tt Dianoga} profile and the corresponding $16^{\rm th}$--$84^{\rm th}$ percentile, while the orange lines and points are for observed profiles for X-COP clusters by \cite{ghirardini.etal.2019}. As for the temperature profiles, we also compare results from the {\tt Dianoga} clusters to the observational measurements of the low-redshift sample by  \cite{mcdonald.etal.2014} (blue solid line and shaded area) and from the Data Release 1 CHEX-MATE sample by \cite{rossetti.etal.2024}. As for the latter, we show with the magenta points the median temperature profiles of the 30 CHEX-MATE clusters. The profiles are normalized according to Eq.~\ref{eq:t500} using the values of $M_{500}$ derived from the {\it Planck} SZ signal, with the method described in \cite{Planck_XXVII}, using the MMF3 algorithm, as discussed in \cite{chex_mate}. As for the temperature profiles by \cite{mcdonald.etal.2014}, they have been measured using {\em Chandra} data for 80 SZ-selected galaxy clusters identified by the {\em South Pole Telescope} (SPT). Of this sample,  we consider here only the low-$z$ subsample that includes the 40 clusters at $z<0.6$. In this paper, temperature profiles have been normalized to the spectroscopic temperature measured within $[0.15-1]R_{500}$. This normalization could in principle differ from that based on Eq. (\ref{eq:t500}). However, comparing the two panels of Figure 18 of \cite{rossetti.etal.2024}, we note that the normalization used in the CHEX-MATE paper, i.e. the spectroscopic temperature measured within the $[0.15-0.75]R_{500}$ radial range, leads to very similar results to those based on $T_{500}$ as in \cite{ghirardini.etal.2019}. For this reason, we assume that it is fair to compare the different temperature profiles shown in Fig. \ref{fig:prof}, although based on different normalizations.

As for the electron number density profiles (upper left of Fig. \ref{fig:prof}) we note that observational results fall within the scatter of the simulated profiles. Still simulations tend to produce slightly flatter profiles than observed at small radii, $\lesssim 0.15R_{500}$. At the same time, temperature profiles within a similar radius tend to be steeper in simulations than in observations (upper right panel of Fig.\ref{fig:prof}), while having comparable slopes beyond $\simeq 0.3 R_{500}$. Combining these differences, we detect a factor-two offset between simulations and observations in the entropy level within the innermost regions (lower left panel of Fig. \ref{fig:prof}). While being characterized by a rather large scatter at small radii, the simulated ICM has on average rather large isentropic cores, with entropy profiles becoming nearly flat below $R\mincir 0.15 R_{500}$. On the other hand, the X-COP entropy profiles from \cite{ghirardini.etal.2019} are steadily decreasing toward the centre, the typical signature of a cool-core structure. An observed ICM at such low entropy is characterized by a short cooling time, that should make it disappear from the hot, X--ray emitting phase in a relatively short time \citep[e.g.][]{voit2005,borgani_kravtsov2011,gaspari.etal.2019}. Therefore, the presence of such cool cores is in general interpreted as the fingerprint of AGN feedback, which provides the energy source compensating on average the radiative losses and maintaining a low-entropy, short cooling-time plasma in the X-ray emitting phase. In this sense, the presence of central isentropic cores signals that the AGN feedback implemented in our simulations is not able to provide a self-regulated compensation of the radiative losses. In fact, a higher entropy level predicted by simulations can be interpreted as either the effect of an exceedingly efficient AGN feedback, that overheats the gas, or as the effect of a too efficient cooling, that selectively removes low-entropy gas, which has a shorter cooling time. Deciding which one of these two explanations is the correct one would require a dedicated analysis, which is beyond the scope of this paper. However, owing to the presence of overmassive BCGs in massive clusters, predicted by simulations (see Fig. \ref{fig:m500_mbcg}), it is plausible that isentropic cores are another manifestation of the inefficient regulation of gas cooling, and of the ensuing star formation, at the centre of massive simulated clusters.

Quite remarkably, the large scatter displayed by simulated entropy profiles at small radii witnesses the sensitivity of this diagnostic to the details of the (thermo-)dynamical state of clusters. At the same time, simulated entropy profiles have a slope that at larger radii has a much lower scatter, is very close to the observed profiles and approaches the slope $K(R)\propto R^{1.1}$ predicted by the self-similar model \citep[e.g.][]{Tozzi2001,voit.etal.2005}. This is in line with the expectations that, outside the core regions, the ICM thermodynamics is mainly determined by gravity-driven hydrodynamical processes and, as such, cluster properties follow a more universal self-similar behaviour \citep[e.g.][]{voit.etal.2005}. In Appendix \ref{sec:app_entr_prof} we will discuss the sensitivity of such profiles to the different implementations of AGN feedback listed in Table \ref{t:models}.

The difficulty in predicting the observed cool-cores is a result shared by other simulations of galaxy clusters presented in the literature and having resolution comparable to our {\tt Dianoga} simulations, such as TNG-Clusters \citep{Nelson.etal.2024,Lehle.etal.2024} and the FABLE simulations \citep{henden.etal.2018}. In particular, we note that the results of thermodynamical profiles of TNG-Clusters presented in \citet{Lehle.etal.2024} are broadly consistent with ours in predicting average entropy profiles that, in the central regions, are shallower than the observed ones, and average temperature profiles that are steeper than observed. 

On the other hand, the disagreement between observed and simulated entropy profiles is at variance with our previous findings presented in \cite{rasia.etal.2015}. In fact, that paper was based on the same set of initial conditions used here, but at a 25 times coarser mass resolution, and using similar implementations for star formation and feedback from SN and AGN. In fact, \cite{rasia.etal.2015} found that simulated entropy profiles show the correct cool core structure, with entropy profiles declining down to the smallest resolved cluster-centric radii. Quite interestingly, similar results on a realistic cool core structure in simulated clusters were also found by \cite{Qingyang.etal.2020} from The Three Hundred simulations, by \cite{gonzalez.etal.2025}, using the Magneticum simulations \citep{dolag.etal.2025} and by \cite{braspenning.etal.2024}, using the Flamingo simulations \citep{schaye.etal.2023}. While covering large cosmological volumes, thus providing a large statistics of simulated clusters, the three aforementioned  simulations have a rather limited resolution, comparable to that of the simulations analysed by \cite{rasia.etal.2015}. Taken together, these results suggest that current implementations of AGN feedback in cosmological simulations could in principle be tuned in such a way to provide realistic cool cores in galaxy clusters. However, as the resolution is increased, self-regulation between cooling and feedback becomes more difficult to obtain, independent of whether AGN feedback is implemented in a purely thermal fashion, like in the simulations presented here and in Magneticum, including also a kinetic feedback mode, as in the TNG-Clusters and Flamingo simulations, or injecting high-entropy bubbles, as in FABLE. As discussed in Sect. \ref{sec:app_entr_prof}, more cool-cored entropy profiles are predicted for the M4-M6 models of Table \ref{t:models}, in which evaporation by AGN feedback of the cold phase of star-forming particles is allowed, at least for the test cases considered (see also Fig. \ref{fig:entr_comp}). This confirms the delicate role played by the interaction between AGN feedback and star formation model. 

We note that, while the electron density and temperature structures combine to provide entropy profiles at variance with observations, their combination produces instead simulated pressure profiles which agree quite nicely with observational results in the core regions. This is in line with the expectations that pressure is a robust indicator of the structure of the gravitational potential, which is in turn mainly determined by the gravitational dynamics and weakly sensitive to the details of the baryonic processes included. In general, the comparison between observational data of ICM thermodynamical profiles and the predictions of our {\tt Dianoga} simulations show a quite good agreement outside the core regions, $R\magcir 0.15 R_{500}$, while the differences in the core regions point to the difficulty of the implemented AGN feedback model in properly regulating the cooling-heating cycle in the central regions of massive clusters. 

\section{Conclusions}
\label{sec:concl}
In this paper, we have presented a comprehensive analysis of a high-resolution version of the {\tt Dianoga} set of cosmological hydrodynamical zoom-in simulations of galaxy clusters and groups, complemented by a control cosmological box of about 50$\,h^{-1}$ comoving Mpc, simulated at the same resolution, that provides a sort of reference "field". The primary purpose of this work was to critically address the persistent challenges that modern cosmological simulations face in simultaneously reproducing the complex cooling-feedback regulation within the core regions of galaxy clusters, the stellar mass function of the accompanying galaxy population, and the thermodynamic properties of the ICM.

To address these issues, we utilized the \OG\ TreePM-SPH code to simulate 28 Lagrangian regions containing a statistically significant sample of 293 halos with a mass larger than $M_{200}\ge 1.5\times 10^{13}\mathrm{M_\odot}$, out of which 74 clusters with $M_{200} \ge 10^{14} \, \mathrm{M_\odot}$ and 23 massive clusters with $M_{200} \ge 10^{15} \, \mathrm{M_\odot}$ (see Table \ref{t:regions}). In our simulations, the initial mass of gas particles is $6.24\times 10^6\,h^{-1}\mathrm{M_\odot}$, while the peak force resolution corresponds to a Plummer-equivalent softening of $250\,h^{-1}$cpc for stellar and BH particles. 

AGN feedback parameters in the reference model are minimally tuned exclusively to reproduce the local  relationship between SMBH mass and host galaxy stellar mass (see Fig. \ref{fig:magorrian}), leaving all group- and cluster-scale properties as genuine predictions of the simulations. To systematically unravel the physical mechanisms driving discrepancies with observations, we investigated a grid of seven distinct implementations of AGN feedback (see Table \ref{t:models}). These variations explored the effect of changing BH feedback efficiency ($\epsilon_f$), Bondi accretion cold/hot boost factors ($\alpha_c, \alpha_h$), the elimination of the Eddington limit to the BH accretion rate, and a modification of the interaction between the sub-resolution AGN and star-formation sectors, according to which AGN feedback energy can evaporate the cold phase within star-forming multi-phase gas particles. Thanks to this combination of resolution, statistics of simulated systems and feedback parameter space explored, our set of simulations provide an extremely useful framework to shed light, through a comparison with observational data, on the impact of the processes driving the baryon cycle and on the delicate interplay between the two essential sub-resolution sectors of BH evolution/AGN feedback and of star formation/SN feedback.

The main results of our analysis can be summarized as follows:
\begin{description}
    \item[{\bf (a)}] Our simulations predict a cluster galaxy stellar mass function (GSMF) that is in general agreement with observations, both in the field and in the environment of groups/clusters, but with a too shallow slope in the high-mass end (see Fig. \ref{fig:cluster_smf}). This witnesses that AGN feedback in our reference M1 model does not regulate star formation in the most massive galaxies to the level required by observations.  Interestingly, our simulations succeed instead in reproducing the steepening of the GSMF at the low-mass end -- $\log(M_*/\mathrm{M_\odot}) \lesssim 10.5$ -- as reported by recent observational investigations \citep{Park.etal.2026}. This indicates that our implementation of stellar feedback via Type-II supernova-driven galactic winds operates at the correct efficiency in low-mass galaxies.
    \item[{\bf (b)}] Consistently with the above results on the high end of the GSMF, in our reference model the stellar masses of BCGs in the most massive clusters, $M_{500}\sim 10^{15}\mathrm{M_\odot}$, are systematically overpredicted, while {\tt Dianoga} results are closer to observations at the scale of groups. Correspondingly, also the stellar mass fraction in the most massive halos exceeds observational measurements by about a factor of three, while being in better agreement for low-mass groups (see Figs. \ref{fig:m500_mbcg} and \ref{fig:starprof}).
    \item[{\bf (c)}] The reference M1 model predicts scaling relations between ICM observables -- the gas mass fraction as a function of halo mass ($
    f_{\rm g}$-$M_{500}$; Fig. \ref{fig:fgas}) and the mass-temperature ($M_g$-$T$; Fig. \ref{fig:mgtsl}) relation -- that are in good agreement with low--redshift observational measurements from group to cluster scales. As for the mass--temperature ($M_{500}$-$T_{500}$, Fig. \ref{fig:mtsl}) relation our {\tt Dianoga} clusters tend to appear slightly more massive than observed, a tension that is alleviated by allowing for a 20 per cent hydrostatic mass bias inherent to X-ray observations. Interestingly, the deviations in gas mass and temperature combine to yield an excellent, robust agreement with the observed $Y_X$--$M_{500}$ relation (Fig. \ref{fig:myx}), reaffirming the role of the X-ray analogue to the Compton-$y$ parameter as a reliable mass proxy which is weakly sensitive to uncertainties in the modelling of baryonic processes.
    \item[{\bf (d)}] A comparison with spatially resolved profiles of ICM properties mostly from the X-COP project \citep[e.g.][]{ghirardini.etal.2019} demonstrates that our simulations accurately capture the thermodynamic structures of clusters beyond the core region ($R \gtrsim 0.15 \, R_{500}$; see Fig.\ref{fig:prof}). In this regime, the simulated profiles of electron number density, temperature, entropy, and pressure converge to observational measurements and to the slopes expected from the self--similar model, thus confirming that the outskirts of massive halos are dominated by gravity-driven processes and shock heating during hierarchical accretion. Within the central cores, the profiles have a diversity of behaviours, ranging from representations of strong ``cool core'' and nearly isentropic ``non-cool core'' systems, mirroring the structural diversity also seen in real clusters \citep[e.g.][]{ghirardini.etal.2019,Sanders.etal.2025}. Still, we note that on average our {\tt Dianoga} clusters are less "cool-cored" than the observed ones. This is consistent with the indication from overmassive BCGs that the heating-cooling cycle predicted by our reference M1 implementation of AGN feedback is not self-regulated at the level implied by observational constraints.
    \item[{\bf (e)}] As shown in Appendix \ref{app:feed}, introducing an explicit sub-resolution mechanism for evaporating the cold phase of star-forming gas particles when they receive AGN thermal energy -- M4, M5, and M6 models -- emerged as an effective solution to bring BCG stellar masses (Fig. \ref{fig:app_mbcg}) and global stellar mass fractions (Fig. \ref{fig:app_fst}) into a closer agreement with observational data, increasing at the same time the cool coreness displayed by the entropy profiles in few test cases (Fig. \ref{fig:entr_comp}). This result demonstrates that the interface between star-formation modules and black hole feedback module is as important as the amount of feedback energy itself made available by accretion onto SMBHs.
\end{description}

In general, it is quite remarkable that our simulations, albeit based on a minimal parameter tuning using only the relationship between SMBH masses and stellar masses of the host galaxies, they are are rather successfull in reproducing a number of observed properties of groups and clusters. Despite such significant improvements, several open issues remain that outline future lines of development. First, while our multi-phase gas evaporation model (M4--M6 models) succeeded in regulating star formation, it remains an isotropic thermal implementation acting on a sub-resolution effective model of star formation. Including an explicit treatment of the kinetic mode of AGN feedback to describe the effect of sub-relativistic jets during periods of low accretion, with directionality explicitly provided by a self-consistent description of the SMBH spin evolution \citep[e.g.][]{Sala.etal.2024} represents the first necessary further step. In fact, including the effect of AGN-driven jets \citep[e.g.][]{Barai.etal.2016} has been shown to be a promising avenue to regulate star formation and produce realistic cool-core structures in the cosmological framework of hierarchical assembly of galaxy clusters \citep[e.g.][]{Weinberger.etal.2026,Rosenberg.etal.2026}. Another direction of improvement should involve the description of the SMBH accretion. While the Bondi criterion is the most common accretion model adopted, its validity relies on assumptions that are not expected to be fulfilled in a cosmological environment. Implementing a model in which the BH gas accretion rate is directly related to the amount of gas cooling within halos should allow to effectively establish a self-regulated balance between heating and cooling in the cores of massive halos \citep[e.g.][]{Gaspari2020}. Finally, the results presented in Appendix \ref{app:feed} from the models including cold gas evaporation highlight that the effect of AGN feedback on star formation sensitively depends on how these two sectors interact with each other, and definitely represents a direction of investigation for forthcoming analyses. In this respect, moving to a star formation model that explicitly treats the mass- and energy-flows between the different ISM phases \citep[e.g.][]{Valentini.etal.2020,Valentini.etal.2023}, also incorporating the effect of \(\mathrm{H}_{2}\) formed on dust grains \citep{Ragone.etal.2024}, would allow a more direct and self-consistent description of the effect of AGN feedback on star formation.

Combining our multi-model simulation approach with available and upcoming observations of galaxy clusters from their infancy at redshift $z\sim 2$--4 to the nearby universe will provide the constraints necessary to shed light on the mechanisms driving the co-evolution of diffuse baryons, galaxies and the SMBH population, ultimately unlocking a fully self-consistent picture of the history of the baryon cycle within the most massive cosmic structures.

\begin{acknowledgements}
 We thank Andrey Kravtsov who provided the stellar density profiles of Fig.~\ref{fig:starprof} and Mariachiara Rossetti who provided the temperature profiles of Fig.~\ref{fig:prof}. We would like to thank Weipeng Lin, Luca Sala, and Heng Yu for stimulating discussions. SB, AD, ER, IM and AS thank the participants of the workshops “The
Most Massive Galaxies and Their Environment Across Cosmic
Times” and “Gravitational Lensing at the turning point:
Current Discoveries, New Frontiers” held at the {\em Sexten Center
for Astrophysics Riccardo Giacconi}, for insightful discussions.
 Simulations have been carried out at the CINECA Supercomputing Center (Bologna, Italy), with computing time assigned through ISCRA-B and ICSC calls, and through CINECA-INAF, CINECA-UNITS agreements. This paper is supported by: the Fondazione ICSC National Recovery and Resilience Plan (PNRR) Project ID CN-00000013 "Italian Research Center on High-Performance Computing, Big Data and Quantum Computing" funded by MUR Missione 4 Componente 2 Investimento 1.4: "Potenziamento strutture di ricerca e creazione di "campioni nazionali di R$\&$S (M4C2-19 )" - Next Generation EU (NGEU); the National Recovery and Resilience Plan (NRRP), Mission 4, Component 2, Investment 1.1, Call for tender No. 1409 published on 14.9.2022 by the Italian Ministry of University and Research (MUR), funded by the European Union – NextGenerationEU–Project Title "Space-based cosmology with Euclid: the role of High-Performance Computing" – CUP J53D23019100001 - Grant Assignment Decree No. 962 adopted on 30/06/2023 by the Italian Ministry of Ministry of University and Research (MUR). SB, MV and AS acknowledge partial financial support from the INFN Indark Grant. SB, AD and MV acknowledge the "Astrofisica Fondamentale 2024" INAF Grant "Black Hole Dynamics and Galaxy Formation from Cosmological Simulations". ER acknowledges financial support from NASA grants 80NSSC25K0006 and 80NSSC25K8009 and from the Chandra Theory Program (TM4-25006X) awarded from the Chandra X-ray Center operated by the Smithsonian Astrophysical Observatory for and on behalf of NASA under contract NAS8-03060.  VB and ER acknowledge partial financial support from the "Astrofisica Fondamentale 2023" INAF Grant “Origins of the ICM metallicity in galaxy clusters". CRF and GLG acknowledge financial support from the European Union's HORIZON-MSCA-2021-SE-01 Research and Innovation Programme under the Marie Sklodowska-Curie grant agreement number 101086388 - Project (LACEGAL). IM acknowledges financial support from the Excellence Cluster ORIGINS2, which is funded by the Deutsche Forschungsgemeinschaft (DFG, German Research Foundation) under Germany’s Excellence Strategy- EXC-2094390783311. KD acknowledges support by the COMPLEX project from the European Research Council (ERC) under the European Union’s Horizon 2020 research and innovation program grant agreement ERC-2019-AdG 882679. This work was supported by the SPACE Centre of Excellence, and funded by the European Union and several partner countries under grant 101093441. SB wishes to thank support by grant NSF PHY-2309135 to the Kavli Institute for Theoretical Physics (KITP). 
\end{acknowledgements}

%

\bibliographystyle{aa} 
\bibliography{biblio}

\begin{appendix}
\nolinenumbers


\section{Accretion history of the overmassive BHs and BCGs}
\label{app:OverBH}

This Appendix is dedicated to tracking the origin of the overmassive BCGs in the M1 model, as shown in Fig. \ref{fig:m500_mbcg}, and of the overmassive BHs appearing in the left panel of Fig. \ref{fig:magorrian}. These two issues are addressed together since, as shown throughout this Appendix, they arise from closely connected processes: the mechanism causing the enhanced SFR of such BCGs also fuels the rapid growth of the BHs. 

\begin{figure}[ht!]
    \centering
    \includegraphics[width=\linewidth]{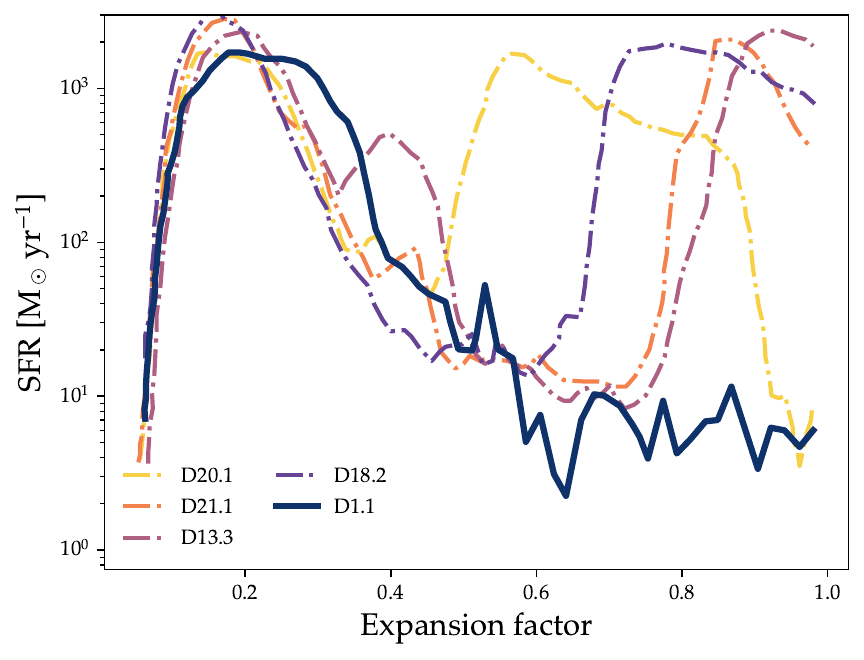}
    \vspace{-0.5truecm}
    \caption{Star-formation rates, as a function of the expansion factor for the four most and overmassive BCGs shown in Figure \ref{fig:m500_mbcg} (dashed lines), and for a "normally" massive BCG (solid line). For each BCG we indicate in the legend the name of the region, as reported in Table \ref{t:regions} and the corresponding mass rank of the halo hosting the BCG. E.g. D21.1 indicates the most massive halo in the D21 region, while D13.3 is for the third most massive halo in the D13 region. 
    }
    \label{fig:sfr_bcg}
\end{figure}

Specifically, we focus on the four overmassive BCGs located in the upper-right region of the left panel of Fig. \ref{fig:magorrian}, corresponding to BCGs in regions D13, D18, D20, and D21 within the M1 model. Figure \ref{fig:sfr_bcg} shows the star formation histories of these BCGs (dash-dotted lines). For comparison, we also show the SFR of a fifth BCG from region D1 in the same model (solid line), which does not exhibit this anomalous overgrowth, to highlight the difference in their evolution. All the SFRs have been computed by identifying the stars contained within the central $50 \, \rm kpc$ at $z=0$, and using their formation redshift to trace back the star formation history. As such, these SFRs do not refer only to the BCG main progenitor. 
All the BCGs show consistently an early peak of star formation at $z \sim 4$. Quite apparently, all the overmassive BCGs experienced recent strong and prolonged bursts of star formation, with rates even exceeding $10^3 \, \mathrm{M_\odot} \, \rm yr^{-1}$. 
On the other hand, the "normal" BCG in the D1 region exhibits a smoothly declining SFR, reaching a value of few $\mathrm{M_\odot} \rm \, yr^{-1}$ at $z=0$. 

\begin{figure*}
    \centering
    \includegraphics[width=0.9\linewidth]{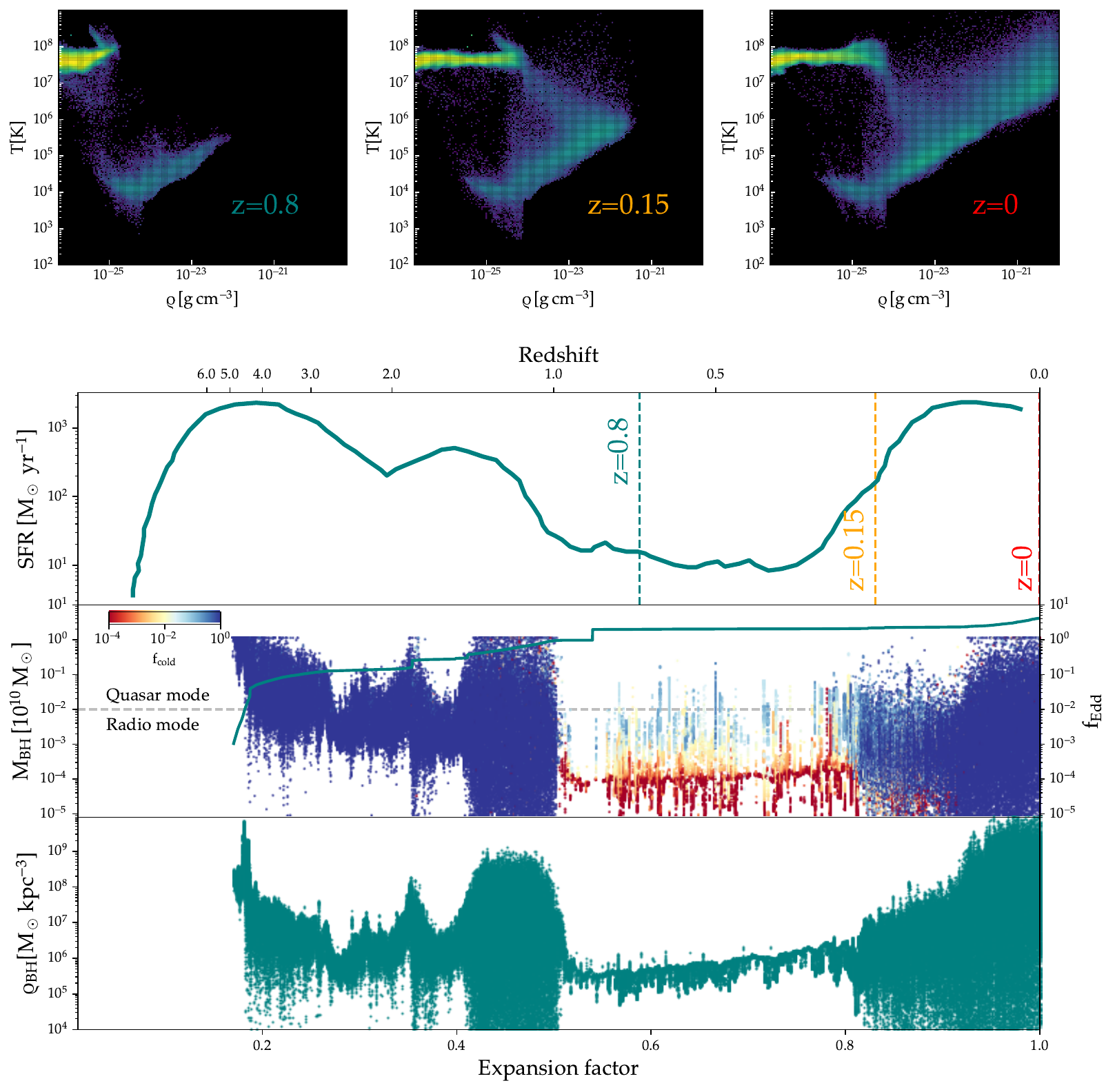}
    \caption{Evolution of BCG SFR and central BH accretion properties for the D13.3 halo. The upper panels show the mass distribution of gas particles in the phase diagram, in which densities are reported in comoving units, for $z=0.8, 0.15, 0$ from left to right. The lower panel reports, as a function of the expansion factor: the SFR of the BCG (top), the BH mass overplotted with the Eddington ratio (central) and the SPH gas density computed at the position of the central BH (bottom). The Eddington ratio is colour-coded according to the cold-gas fraction $f_{\rm cold} = \rho_{\rm cold}/\rho_{\rm BH}$, with blue indicating cold-gas-dominated accretion and red indicating hot-gas-dominated regimes.}
    \label{fig:D13_accretion_and_SFR}
\end{figure*}

\begin{figure*}
    \centering
    \includegraphics[width=0.9\linewidth]{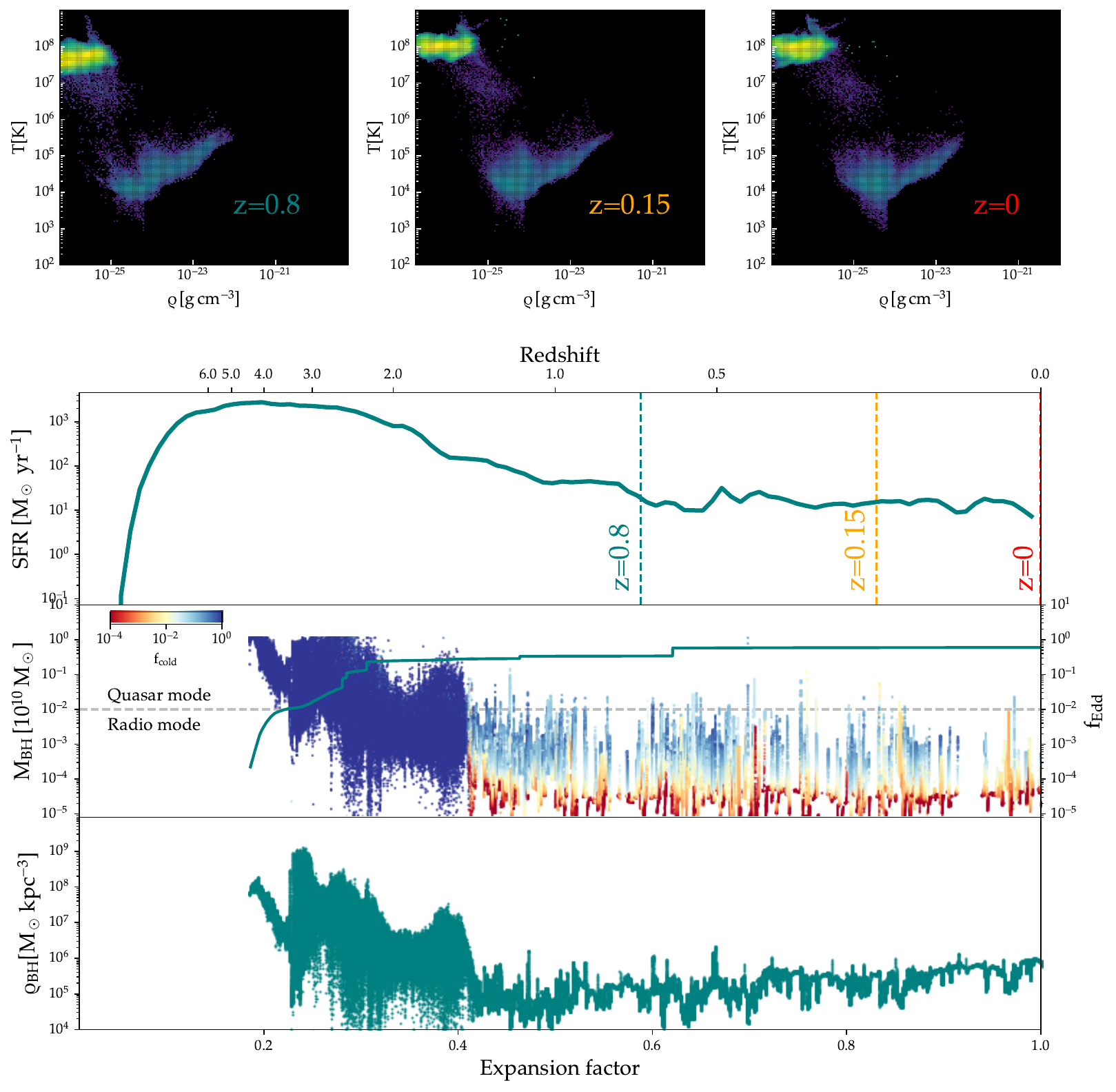}
    \caption{Same as Fig. \ref{fig:D13_accretion_and_SFR} but for the main halo of the D1 simulation. }
    \label{fig:D1_accretion_and_SFR}
\end{figure*}

In order to track the origin of such  behaviours, we show in Fig. \ref{fig:D13_accretion_and_SFR} and \ref{fig:D1_accretion_and_SFR} the coupled evolution of the SFR and of the central SMBH at $z=0$ for the pathological D13.3 halo (Fig. \ref{fig:D13_accretion_and_SFR}) and the "normal" D1.1 case (Fig. \ref{fig:D1_accretion_and_SFR}) along with the phase diagrams in the central densest part of the two halos, at three redshifts. The lower part of each figure shows (from top to bottom): the evolution of the SFR computed using the same method as in Fig. \ref{fig:sfr_bcg}, the BH mass evolution overplotted to the Eddington ratio $f_{\rm Edd} = \dot{M}_{\rm BH}/\dot{M}_{\rm Edd}$, and the gas density at the BH position as a function of the scale factor. Furthermore, the Eddington ratio is colour-coded according to the fraction of cold gas contributing to the density at the BH position, where blue tones indicate a cold-gas dominated accretion, while red tones correspond to hot-gas dominated state. In the upper part of each figure, we report the mass distribution in the density-temperature plane (the so-called "phase" diagram) of the particles lying within a radius of $500 \,  \rm ckpc$ from the halo centre. We focus on particles with physical densities $\rho > 10^{-27} \,\rm g \, cm^{-3}$, close to the star-formation density threshold, in order to highlight the area populated by particles that fuel star formation. We report the phase diagrams for $z=0.8, 0.15, 0$ corresponding to the epochs when, in the D13 region of Fig. \ref{fig:D13_accretion_and_SFR}, the SFR is smoothly decreasing ($z=0.8$), subsequently rises ($z=0.15$) and finally remains sustained until the present epoch ($z=0$).

The early star formation history of both clusters, down to $z\simeq2$, is characterized by a peak of the SFR, which falls within a phase of prolonged cold-gas-dominated accretion onto the central BH. At a lower redshift, $z\simeq 1.0$--1.5, the central BH enters in a more quiescent accretion regime in which the Eddington ratio falls below the radio-mode threshold of 0.01, and the accretion becomes predominantly dominated by hot gas. 
By $z=0.8$ the progenitors of both BCGs exhibit SFR values of $\sim 10 \, \mathrm{M_\odot} \rm \, yr^{-1}$ sustained by particles in the star forming branch at a temperature of a few times $10^{5} \, \rm K$ and density $ \sim 10^{-22} \rm \, g \, cm^{-3}$. 

At lower redshift, the evolutions of the two systems diverge. As for D1, shown in Fig. \ref{fig:D1_accretion_and_SFR}, it remains largely unchanged: the SFR steadily decreases down to $z=0$ and the BH mass growth is quite modest. The Eddington ratio remains mainly below $10^{-5}$, with few accretion spikes triggered by episodic infall of cold gas, which are rapidly suppressed by the AGN feedback response.

On the contrary, the evolution of D13.3 in Fig. \ref{fig:D13_accretion_and_SFR} exhibits a simultaneous rejuvenation both of the BCG star formation and of the central BH activity. As shown in the bottom panel of Fig. \ref{fig:D13_accretion_and_SFR}, the total gas density surrounding the central BH starts rising below $z\simeq 0.8$, indicating an inflow of gas toward the cluster core. By $z\simeq 0.15$ the central panel reveals that this gas inflow powers the Eddington ratio to values $>10^{-2} \, M_{\rm Edd}$ with the accretion being dominated by the cold gas. The corresponding phase diagram at $z=0.15$ shows the development of an isothermal core at a temperature $\simeq 3\times 10^7\, K $ whose density eventually reaches the value of the star-formation threshold. This mechanism triggers a "rain" of particles that rapidly reaches the star-forming branch, thus sustaining the extremely high SFR that characterizes this system at low redshift. 

When this violent inflow reaches the BH smoothing length, the accretion rate raises up to the Eddington limit, causing a rapid increase of the BH mass. Despite the strongly enhanced accretion activity, the AGN feedback ultimately fails to quench this excess of gas cooling and growth. The combined effect of the dense shock-heated gas cooling onto the star-forming branch, which replenishes the star-forming branch of the phase diagram, and the BH feedback acting on multi-phase gas particles produces a thickening of the star-forming branch, which extends up to temperatures of $T=10^{8} \, \rm K$ and densities $\rho > 10^{-21} \, \rm g \, cm^{-3}$.

The process illustrated here for the D13.3 cluster is the same responsible for the excessive growth of both BH masses and stellar masses of the host BCGs in the sample of overmassive BH systems shown in Fig. \ref{fig:magorrian}.
In summary, these pathological systems emerge whenever the density of the hot ICM phase exceeds the star-formation density threshold. For the case of D13.3 we verified that this density increase is triggered by a relatively minor merger penetrating to the core regions. Whenever this happens, the cooling time of the gas particles belonging to the star-forming branch becomes shorter than the heating timescale associated with AGN feedback. As a consequence, gas continues to cool and condense, thus sustaining both star formation and BH accretion over extended periods of time. 

This behaviour highlights an intrinsic limitation of the purely thermal AGN feedback implemented here, which is unable to efficiently prevent such catastrophic, albeit episodic, cooling/accretion events. On the other hand, these results also highlight the delicate interaction between AGN feedback and the SH03 sub-resolution model of star formation implemented in our simulations, which is based on the presence of multi-phase ISM-like particles. This points to the need for a different AGN feedback channel which is capable of directly removing or disrupting the cold gas reservoir in such multi-phase particles.
Our results show that the co-evolution of central BHs and of their host BCGs can follow pathways that depart significantly from the simple picture of AGN-driven quenching of star formation, with their growth becoming strongly intertwined through the combined action of numerical and physical processes regulating gas cooling, star formation, and feedback.

\section{Further comparison with observed BCG stellar masses}
\label{app:bcg}
\begin{figure}
    \centering
\includegraphics[width=1.\linewidth]{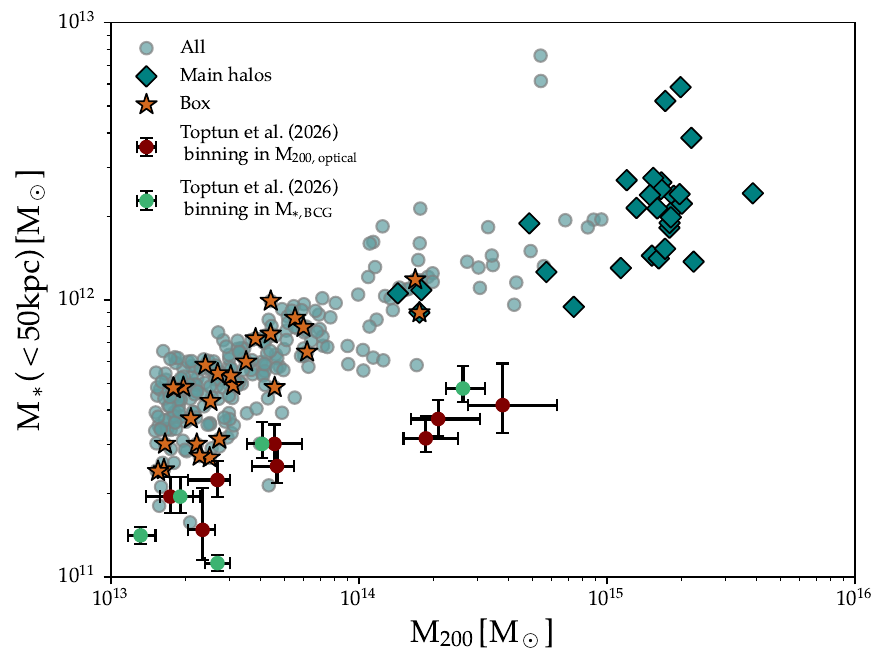}
\vspace{-.5truecm}
    \caption{The same as the left panel of Fig. \protect\ref{fig:m500_mbcg}, but for $M_{200}$ halo masses. Observational data points are now taken \protect\cite{Toptun.etal.2026}, with red and green circles with errorbars corresponding to their analysis based on binning data according the values of $M_{200}$ and of BCG stellar masses, respectively.}
      \label{fig:m200_mbcg}
\end{figure}

In this appendix we extend the comparison of the BCG stellar masses predicted by the M1 model with observed BCG stellar masses, which have been already discussed in Sect. \ref{s:bcgs}. Predictions from the {\tt Dianoga} simulations are compared here with the recent results by \cite{Toptun.etal.2026}, who measured the relationship between BCG stellar masses within a 50 kpc aperture and cluster masses measured within $R_{200}$, instead of $R_{500}$ as shown in Fig. \ref{fig:m500_mbcg}. In their analysis, these authors measure BCG stellar masses using SDSS photometry and halo masses in the range $(10^{12}$--$10^{15})\,\mathrm{M_\odot}$  from an X--ray stacking analysis of eROSITA eRASS1 data. The results shown in Fig. \ref{fig:m200_mbcg} largely confirm that our simulations for the M1 model predict BCGs that are overmassive at the scale of clusters, while they tend to approach the observed ones at the scale of smaller groups. In Appendix \ref{sec:app_fst} we will discuss the effect of changing the implementation of AGN feedback on the BCG stellar masses. 

\section{Effect of changing the AGN feedback model}
\label{app:feed}
In this Appendix we compare the results of the reference M1 model with those from the other models described in Table \ref{t:models}. This provides us with detailed information on the sensitivity of the predictions of our simulations on the implementation of the AGN feedback model. 

One of the variants reported in Table \ref{t:models} concerns the possibility that this feedback causes an evaporation of the cold component of the multi-phase gas particles. We describe here below the implementation of this feature, which is included in the M4, M5, and M6 models \citep[Table \ref{t:models}; see also][]{Damiano.etal.2026}. In the spirit of the SH03 effective model of star formation, multi-phase gas particles that are heated by any source of energy feedback, decay on the equation of state of the effective model over a short time-scale (see Eq. 12 of SH03). Since these particles typically remain at a density exceeding the threshold for star formation after an heating episode, the AGN feedback acting on them is generally ineffective in altering their internal energy over a long time-scale and, therefore, in promoting them to be part of an outflow. To overcome this limitation, we implement in our simulations a scheme to evaporate the cold phase in star forming particles, thus forcing them to exit from the multi-phase stage. If the energy that one such particle receives is sufficient to heat the cold phase, assumed in the SH03 model to be at a temperature of $10^3$K, to the SPH temperature of the particle, then the particle is assumed to exit from the multi-phase stage. However, in the next time-step, the density of such a particle is likely to be still exceeding the star formation threshold, thus implying that it re-enters star formation regime right after. To prevent this from happening, we introduce an additional criterion based on temperature, for such particles to re-enter in the multi-phase stage. We define $u$ to be the specific internal energy of a particle, whose density exceeds the star formation threshold, and $u_{\rm eff}$ the specific internal energy pertaining to such a particle according to the SH03 effective model (see Eq. 19 in their paper). We assume that this particle is prevented from becoming star forming as long as $u> f_{\rm sf}\,u_{\rm eff}$. In this relation, $f_{\rm sf}\magcir 1$ is a fudge factor that specifies how close to the effective equation of state a gas particle must be by cooling to become again star forming. For the M4-M6 simulations presented here, we assume $f_{\rm sf}=1.3$, and we verified that results change only marginally by varying this factor in the range 1--3.

In the tests presented in this Appendix, we concentrate on those observables for which the M1 model is shown to be more in tension with observational data, namely overmassive BH (Appendix \ref{sec:app_MBHMST}), the stellar mass fraction and BCG stellar masses (Appendix \ref{sec:app_fst}) and entropy profiles (Appendix \ref{sec:app_entr_prof}). We verified that using models other than M1 leads only to marginal changes in the ICM scaling relations presented in Sect. \ref{sec:scal}.

\subsection{$M_{BH}$--$M_*$ relation}
\label{sec:app_MBHMST}
\begin{figure*}
    \centering
\includegraphics[width=1.\linewidth]{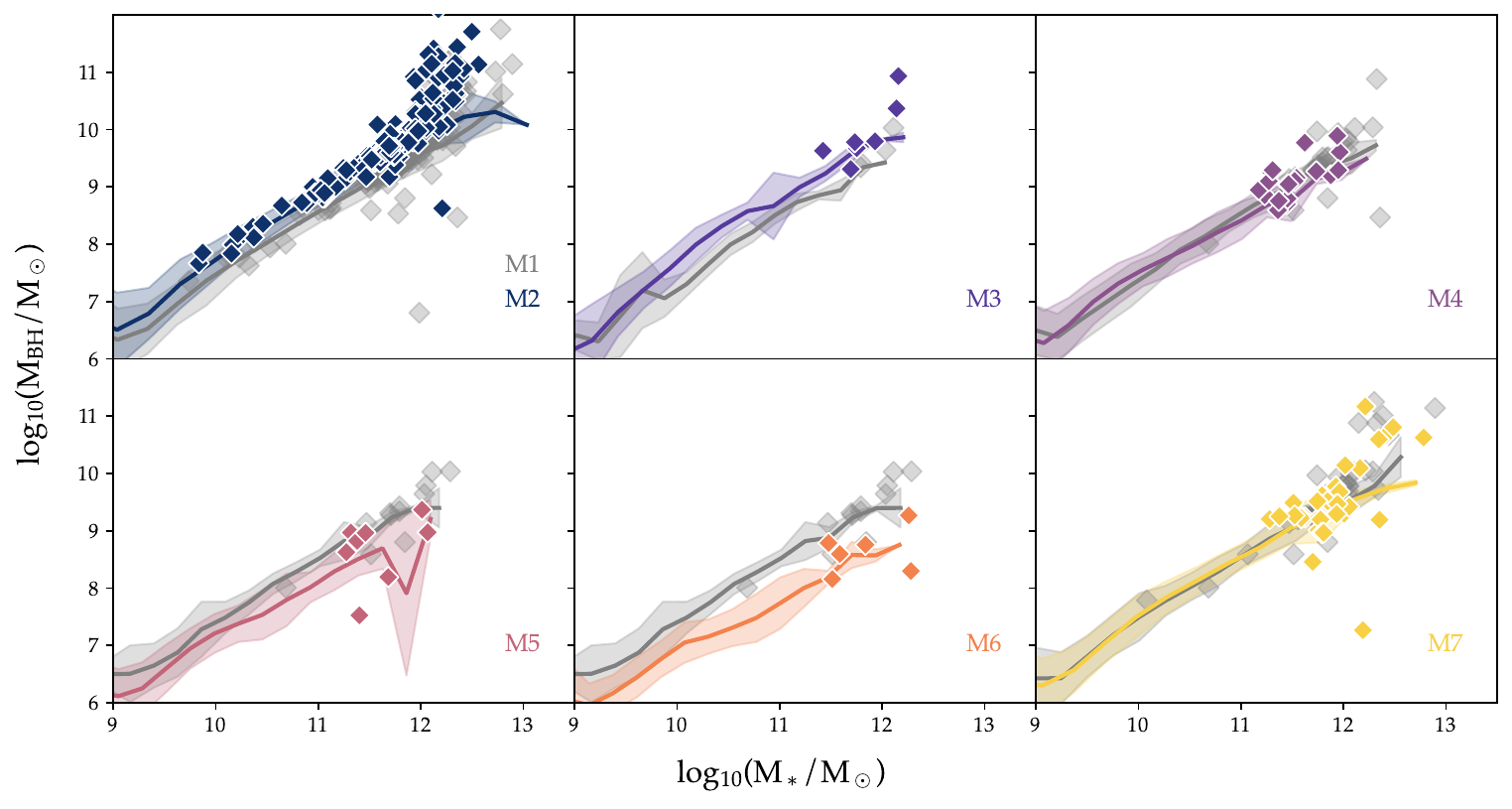}
\vspace{-.5truecm}
    \caption{The relationship between the BH masses and stellar masses of the host galaxies for the AGN feedback models. Each panel shows the comparison between the M1 model and the corresponding model reported in the legend. The meaning of shaded area and diamonds are the same as in Fig.\protect\ref{fig:magorrian}. Each comparison is presented for the {\tt Dianoga} regions in common between M1 (light gray) and each of the other six models.}
      \label{fig:mag_comp}
\end{figure*}
We show in Figure \ref{fig:mag_comp} the results on the relationship between BH masses and stellar masses of the host galaxies for all the seven simulated models. In each panel the results from M1 are compared with those from each of the other six models, only for the {\tt Dianoga} regions that are in common between M1 and each of the other models (see Table \ref{t:regions}). 

As expected, decreasing the feedback efficiency from $\epsilon_f=0.1$ (M1) to 0.05 (M2) makes the masses of the most massive BCGs to further increase, and the overall normalization of the $M_{BH}$--$M_*$ scaling relation to increase. As for this comparison, we refer to the description of Fig. \ref{fig:magorrian} presented in the main text. As for the comparison with the M3 model, we note that decreasing the values of the $\alpha_c$ and $\alpha_h$ boost factors in the expression of the Bondi accretion rate contributed by cold and hot particles (see Eq. \ref{eq:bondi}) does not lead to significant differences. If any, we note that despite the lower accretion rate, the final normalization of the scaling relation, as well as the masses of the most massive BHs somewhat increases. This is interpreted as the effect of a less efficient feedback in the pristine phases of BH accretion, which causes a larger amount of high-density gas to remain available for a more efficient accretion during later stages. This result witnesses the non-linear interplay between accretion efficiency, feedback cycle and the underlying model of star formation. 

The results for the M4, M5 and M6 models show the effect of introducing the evaporation of multi-phase gas particles, according to the scheme described above. As for the M5 model, it is identical to M1, except for the introduction of the evaporation of the cold phase. This feature leads to a lower normalization of the scaling relation, as well as a reduction of the stellar masses of the most massive BCGs and of the hosted SMBHs. This is in line with our analysis of the origin of the phase of exceedingly intense BH accretion discussed in the previous Appendix \ref{app:OverBH}: allowing high-density multi-phase gas particles to leave the star forming phase prevents such particles from contributing to the sporadic runaway accretion episodes, which are at the origin of over-massive BHs. At the same time, the removal of such particles from the star-forming phase makes AGN feedback more efficient in regulating star formation in the BCGs, thereby decreasing their masses. Decreasing the feedback efficiency to $\epsilon_f=0.05$ in the M4 model makes the normalization of the $M_{BH}$--$M_*$ scaling relation nearly identical to that of the M1 model, while preventing overmassive BHs and correspondingly reducing BCG stellar masses. On the contrary, further increasing the efficiency to $\epsilon_f=0.2$ produces a too low normalization of this scaling relation. 

As a final test, in the M7 model we remove the Eddington limit for the BH accretion rate. Quite apparently, the results are virtually unchanged with respect to the M1 model. The reason for this is that, at the resolution of our simulations, episodes of super-Eddington accretion are quite rare and take place at relatively high redshift (see Fig. \ref{fig:D1_accretion_and_SFR}). While Fig. \ref{fig:D13_accretion_and_SFR} shows evidence of Eddington limited accretion even down to $z=0$, we remind that this has to be considered as a "pathological" case that is eventually prevented by allowing AGN feedback to evaporate the ISM in star forming particles. 

\subsection{Stellar content}
\label{sec:app_fst}
\begin{figure*}
    \centering
\includegraphics[width=1.\linewidth]{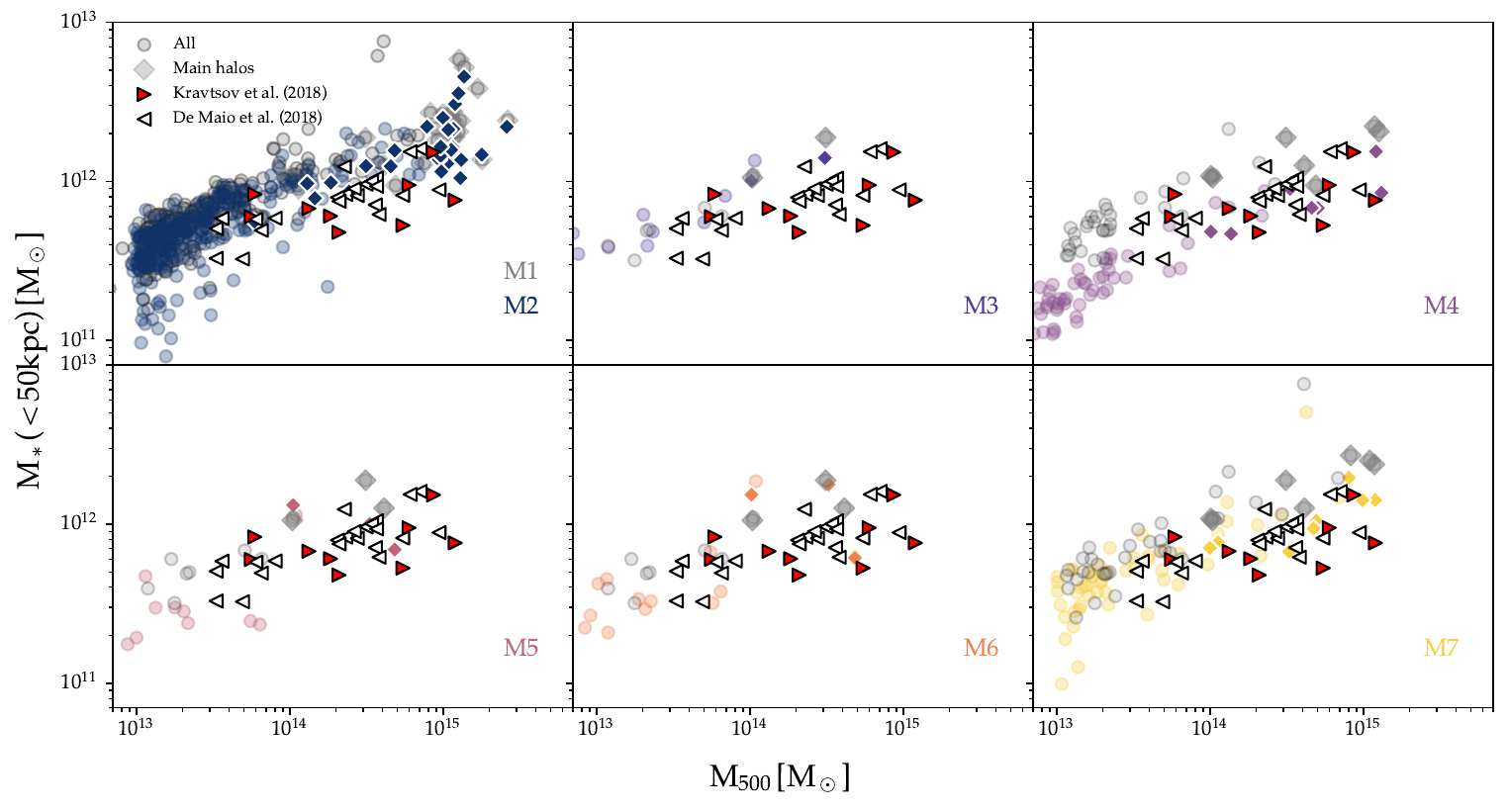}
\vspace{-.5truecm}
    \caption{The same as Fig.\protect\ref{fig:mag_comp} but for the relationship between BCG stellar masses and total cluster mass within $R_{500}$. For reference, we also show the observational results from  \protect\cite{DeMaio2018} and \protect\cite{kravtsov.etal.2018}.}
      \label{fig:app_mbcg}
\end{figure*}

\begin{figure*}
    \centering
\includegraphics[width=1.\linewidth]{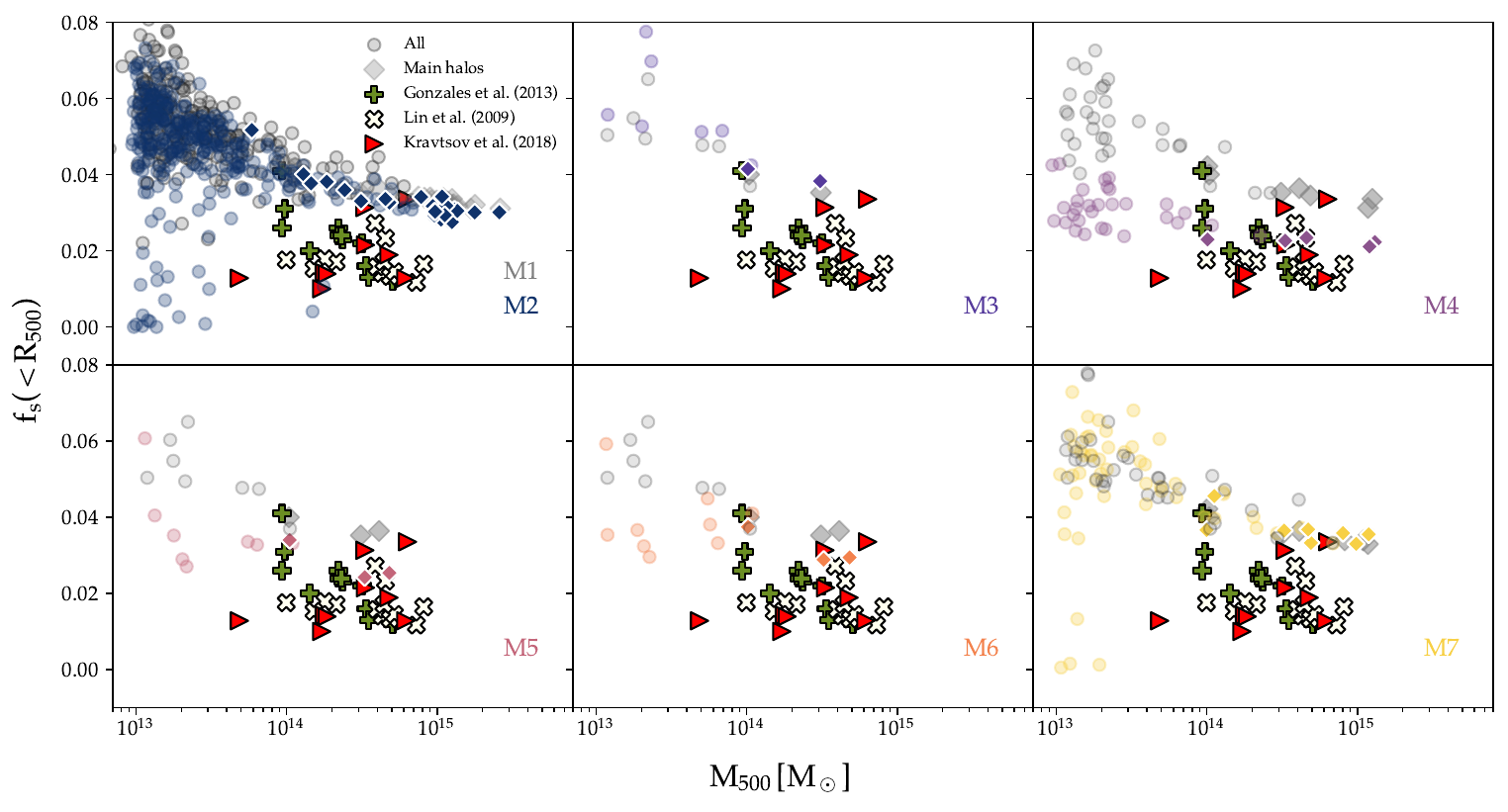}
\vspace{-.5truecm}
    \caption{The same as Fig.\protect\ref{fig:mag_comp} but for the relationship between stellar mass fraction and total cluster mass within $R_{500}$. For reference, we also show the observational results from \protect\cite{lin.etal.2003}, \protect\cite{gonzalez.etal.2013} and \protect\cite{kravtsov.etal.2018}.}
      \label{fig:app_fst}
\end{figure*}

We show in Figures \ref{fig:app_mbcg}  and \ref{fig:app_fst} the effect of changing the AGN feedback model on the BCG stellar masses and on the stellar mass fraction, respectively. As discussed in Sect. \ref{sec:star}, for these two observables the M1 model has been shown to be in significant tension with observations (see  Fig.\ref{fig:m500_mbcg}). 

Consistently with the results of the previous Appendix \ref{app:OverBH}, changing feedback efficiency in M2 or decreasing the accretion boost factors in M3 does not lead to any significant change in both the BCG stellar masses and in the overall stellar mass fraction. As for allowing super-Eddington accretion in M7, it does not produce a sizeable change in the overall stellar mass fraction. On the other hand, it leads to a decrease of the BCG stellar masses thus partially alleviating the tension with observational data. We interpret this result as due to the increased AGN feedback associated to the super-Eddington accretion phases experienced by the progenitors of the BHs hosted in the BCGs, which leads to an improved regulation of star formation. This result highlights the need to include in simulations an improved model for BH accretion that self-consistently include the possibility of super-Eddington accretion phases not only for its impact on the early phases of BH accretion \citep[e.g.][]{Massonneau.etal.2023}, but also to regulate the subsequent star formation in their host galaxies. 

As for the models including evaporation of the cold phase in star-forming particles (models M4 to M6), they are confirmed to be effective in reducing both the BCG stellar masses (Fig. \ref{fig:app_mbcg}) and the stellar mass fraction (Fig. \ref{fig:app_fst}) to the observed level, almost irrespectively of the assumed  feedback factor $\epsilon_{\rm f}$. This result confirms that the implementation of the interplay between star-formation sector and BH sector is essential in determining the impact of AGN feedback on key observables of galaxy clusters and groups.     

\subsection{Entropy profiles}
\label{sec:app_entr_prof}
\begin{figure*}
    \centering
\includegraphics[width=1.\linewidth]{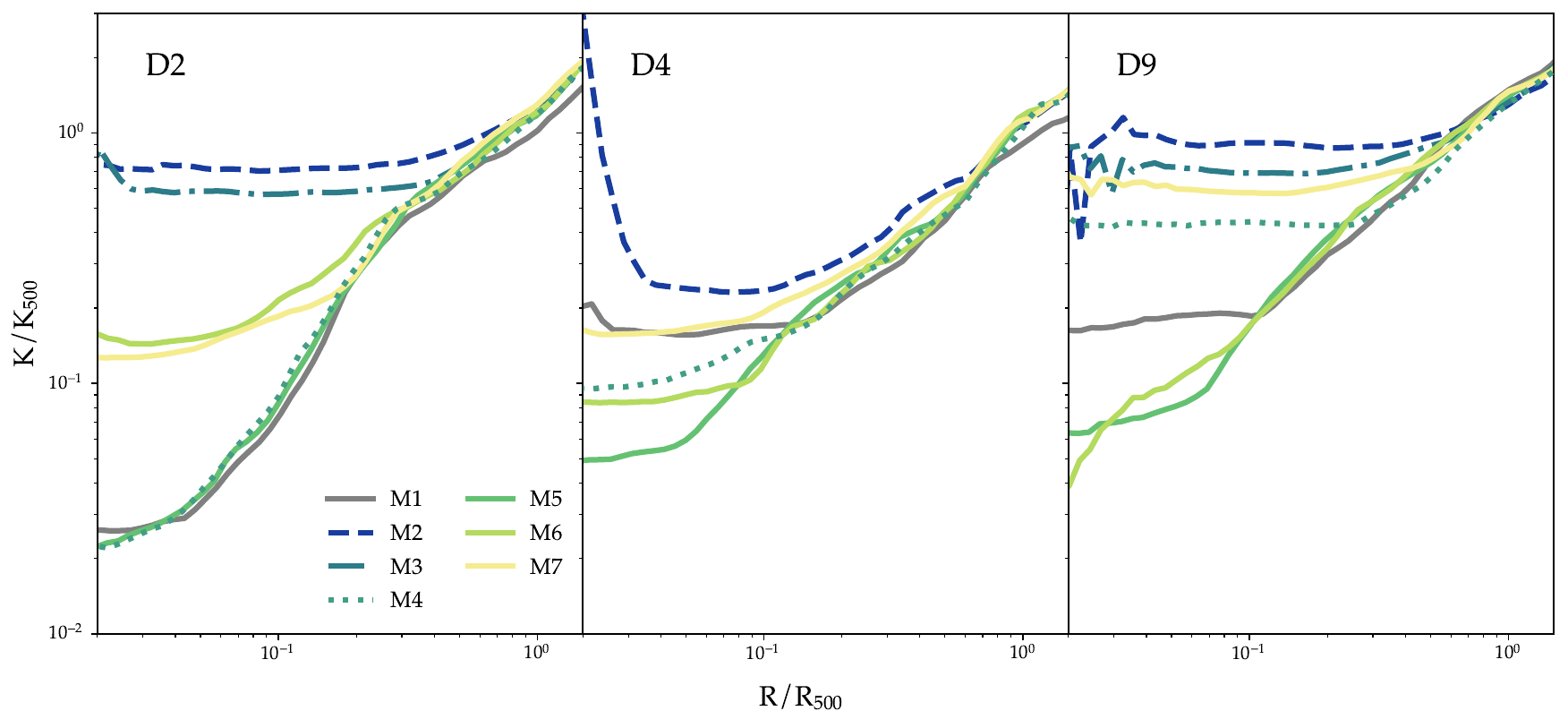}
\vspace{-.5truecm}
    \caption{Comparison of the entropy profiles for the simulated models described in Table \protect\ref{t:models}, for the main halos of the D2 (left), D4 (middle) and D9 (right) regions.}
      \label{fig:entr_comp}
\end{figure*}

We show in Figure \ref{fig:entr_comp} the entropy profiles of the main halos of the D2, D4 and D9 {\tt Dianoga} regions. These three regions have been chosen since they have been simulated with the largest number of different implementations of the AGN feedback model (see Table \ref{t:regions}). At the same time, they have rather different behaviours in the reference model. In fact, the main halo of D2 in the M1 version has an entropy profile decreasing down to the smallest  radii, characteristic of a strong cool core, while D4 and D9 both show nearly isentropic cores for $R\mincir 0.1R_{500}$.

Quite remarkably, entropy profiles in the outskirts, $R\magcir R_{500}$, are all quite similar, independently of the AGN feedback model implemented. This is consistent with the picture that gravity-driven processes of accretion shocks dominate the ICM thermodynamics in this regime. On the other hand, at smaller radii each cluster displays a range of behaviours, depending on the details of the AGN feedback implementations. For instance, in the D2 cluster we note that decreasing the feedback efficiency from $\epsilon_{\rm f}=0.1$ (M1) to 0.05 (M2) makes the object transitioning from a cool-core behaviour to markedly non cool-core one, characterized by a large isentropic core. This witnesses once more how delicate is the interplay between AGN feedback and radiative cooling in establishing the ICM thermodynamical properties in the core regions of galaxy clusters, and how difficult is to establish a self-regulated cooling-feedback loop which prevents the onset of either runaway cooling or catastrophic heating. 

As a general trend in the entropy profiles of these three clusters, we note that the M2 model produces the highest entropy levels in the core regions. Owing to the lower feedback efficiency of this model, we argue that this higher entropy is the result of a selective removal of low entropy gas by cooling, rather than of a strong heating. The exception is possibly represented by the innermost region of D4. In this case, the presence of a steep entropy rise signals the presence of a recent episodic event of AGN heating. The resulting negative entropy gradient is characteristic of a convectively unstable intra-cluster atmosphere, that would produce the displacement of the heated gas from the core regions by buoyancy.  

As for the M4-M6 models, which include the evaporation of the cold phase in star-forming particles, we note that they produce on average rather low entropy levels in core regions, with profiles that are typical of either strong or moderate cool-core clusters. This confirms that modelling the interaction of AGN feedback energy with the sub-resolution model of star formation represents a crucial aspect to establishing the cooling-heating balance. 

In summary, entropy profiles are confirmed to provide a sensitive diagnostic of the thermal status of the ICM. Their diversity, when considering both different objects and different feedback implementations for the same object, highlights that entropy is in fact a tracer of the instantaneous action of feedback on gas in the core regions of clusters and groups.

As a final note, although the M4–M6 models, which include the evaporation of the cold phase of multi-phase gas particles, are shown above to bring BCG stellar masses, stellar mass fractions and entropy profiles into closer agreement with observations, we regard this implementation as still experimental at this stage. The evaporation scheme introduced here has not yet undergone the same extensive testing and systematic calibration applied to the reference AGN feedback model, and its robustness across the full statistical sample of Dianoga clusters remains to be established. We defer a more thorough validation and calibration of the cold-phase evaporation scheme to future work.

\FloatBarrier 
\clearpage

\end{appendix}
\end{document}